\documentclass[namedreferences]{SolarPhysics}
\usepackage[optionalrh]{spr-sola-addons} 
\usepackage{graphicx}                    
\usepackage{amssymb}                     
\usepackage{color}                       
\usepackage{url}                         
\usepackage{textcomp}
\usepackage{hyperref}
\graphicspath{{./}}

\hypersetup{
    unicode=false,          
    pdftoolbar=true,        
    pdfmenubar=true,        
    pdffitwindow=false,     
    pdfstartview={FitH},    
    pdftitle={The Flare-energy Distributions Generated by Kink-unstable
Ensembles of Zero-net-current Coronal Loops},    
    pdfauthor={M. R. Bareford} {P. K. Browning} {R. A. M. Van der Linden},     
    pdfsubject={The Energy Distributions Generated by Zero-net-current Coronal Loops}, 
    pdfcreator={M. R. Bareford}, 
    pdfproducer={M. R. Bareford}, 
    pdfkeywords={Instabilities} {Magnetic fields} {Magnetic reconnection} {Magnetohydrodynamics (MHD)} {Plasmas} {Sun: corona}, 
    pdfnewwindow=true,      
    colorlinks=true,        
    linkcolor=blue,         
    citecolor=blue,         
    filecolor=blue,         
    urlcolor=blue           
}

\begin{document}

\begin{article}

\begin{opening}

\title{The Flare-energy Distributions Generated by Kink-unstable
Ensembles of Zero-net-current Coronal Loops}

\author{M. R.~\surname{Bareford}$^{1}$\sep P. K.~\surname{Browning}$^{1}$\\
R. A. M.~\surname{Van der Linden}$^{2}$}

\runningauthor{M. R. Bareford, P. K. Browning, and R. A. M. Van der Linden}
\runningtitle{The Energy Distributions Generated by Zero-net-current Coronal Loops}

\institute{$^{1}$ Jodrell Bank Centre for Astrophysics, Alan Turing Building, School of Physics and Astronomy, The University of Manchester, Oxford Road, Manchester M13 9PL, U.K.\\
email: \url{michael.bareford@postgrad.manchester.ac.uk}\\
email: \url{p.browning@manchester.ac.uk}\\ 
$^{2}$ SIDC, Royal Observatory of Belgium, Ringlaan 3, B-1180 Brussels, Belgium\\
email: \url{ronald.vanderlinden@oma.be}}

\begin{abstract}
It has been proposed that the million degree temperature of the corona is due to the combined effect of barely-detectable energy releases, so called nanoflares, that occur throughout the solar atmosphere. Alas, the nanoflare density and brightness implied by this hypothesis means that conclusive verification is beyond present observational abilities. Nevertheless, we investigate the plausibility of the nanoflare hypothesis by constructing a magnetohydrodynamic (MHD) model that can derive the energy of a nanoflare from the nature of an ideal kink instability. The set of energy-releasing instabilities is captured by an instability threshold for linear kink modes. Each point on the threshold is associated with a unique energy release and so we can predict a distribution of nanoflare energies. When the linear instability threshold is crossed, the instability enters a nonlinear phase as it is driven by current sheet reconnection. As the ensuing flare erupts and declines, the field transitions to a lower energy state, which is modelled by relaxation theory, i.e., helicity is conserved and the ratio of current to field becomes invariant within the loop. We apply the model so that all the loops within an ensemble achieve instability followed by energy-releasing relaxation. The result is a nanoflare energy distribution. Furthermore, we produce different distributions by varying the loop aspect ratio, the nature of the path to instability taken by each loop and also the level of radial expansion that may accompany loop relaxation. The heating rate obtained is just sufficient for coronal heating. In addition, we also show that kink instability cannot be associated with a critical magnetic twist value for every point along the instability threshold.
\end{abstract}

%

\end{opening}

%
\section{Introduction}
\label{sec:Introduction} 
The theory of nanoflare coronal heating (Parker, \citeyear{Parker1988}) postulates that small flares are sufficiently numerous to maintain a coronal temperature of millions of degrees. Observational studies have so far been unable to show conclusively whether nanoflares occur frequently enough to heat the corona (Krucker and Benz, \citeyear{Krucker1998}; Parnell and Jupp, \citeyear{Parnell2000}; Aschwanden and Parnell, \citeyear{Aschwanden2002}; Parnell, \citeyear{Parnell2004}): the smaller a flare the harder it is to distinguish from the coronal background. Nevertheless, we propose a coronal loop model, the purpose of which is to test the viability of the nanoflare theory with respect to coronal heating. 

Random convective motions at the photosphere increase the energy of the magnetic fields that define coronal loops (plasma beta, $\beta$\,$\approx$\,0.01). The magnetic field is repeatedly twisted by such motions; coronal conductivities are so large that the magnetic flux and plasma can be regarded as being frozen together. We assume that the kinetic energy imparted by the photosphere is dissipated via Direct Current heating: the timescale for photospheric turbulence is long compared to the Alfv\'en time (Klimchuk, \citeyear{Klimchuk2006}). Hence, the loop can be represented as moving through a series of force-free states: $\nabla\times\vec{B}\,=\,\alpha (\vec{r})\vec{B}$, where $\alpha\,=\,(\mu_{0}\vec{j}\cdot\vec{B})$/$(|\vec{B}|^{2})$ is the ratio of current density to magnetic field and \textit{$\vec{r}$} is a position vector (Woltjer, \citeyear{Woltjer1958}). Magnetic stresses build within the loop until an instability is reached or dissipation is otherwise triggered.

Magnetic reconnection is thought to be the mechanism by which the excess magnetic energy is converted into heat. Observations of large-scale flares have revealed stong evidence for such a process (Fletcher, \citeyear{Fletcher2009}; Qiu, \citeyear{Qiu2009}). In addition, several 3D MHD models have shown that current sheets form during the nonlinear phase of an ideal kink instability (Baty and Heyvaerts, \citeyear{Baty1996}; Velli, Lionello, and Einaudi, \citeyear{Velli1997}; Arber, Longbottom, and Van der Linden, \citeyear{Arber1999}; Baty, \citeyear{Baty2000}). The kink instability gives rise to helical current sheets that become the site of Ohmic dissipation, i.e., a heating event. These models are restricted to a narrow range of initial loop configurations. Expanding this range so that one could determine the relationship between say, the initial $\alpha$-profile and the resulting energy release would be too computationally expensive. However, the energy release can be calculated without recourse to following the complex dynamics of the reconnection process (Heyvaerts and Priest, \citeyear{Heyvaerts1984}).

When a magnetic field becomes unstable, relaxation theory predicts the field will transition to the lowest energy state that conserves total magnetic axial flux and \textit{global} magnetic helicity (Taylor, \citeyear{Taylor1974}, \citeyear{Taylor1986}). The minimum energy (or relaxed) state is the well-known constant $\alpha$ or linear force-free field, $\nabla\times\vec{B}\,=\,\alpha\vec{B}$. The helicity (\textit{K}) measures the degree to which the magnetic field is linked with itself (Berger, \citeyear{Berger1999}). However, the relative helicity (Berger and Field, \citeyear{BergerField1984}; Finn and Antonsen, \citeyear{Finn1985}) is more useful since it is gauge invariant:
\begin{eqnarray}
  \label{relative_helicity}
  K & = & \int_V (\vec{A}+\vec{A'})\cdot (\vec{B}-\vec{B\,'})\,\,dV,
\end{eqnarray}
where \textit{$\vec{A}$} is the magnetic potential, \textit{$\vec{B}'$} is the potential field with the same boundary conditions and \textit{$\vec{A}'$} is the corresponding vector potential.

Helicity conservation is not absolute. During relaxation, helicity is still subject to global resistive diffusion, but the change in helicity is negligible when compared to the drop in magnetic energy, so long as dissipation predominantly occurs within thin current sheets. The rates of dissipation for helicity and magnetic energy (\textit{W}) are
\begin{eqnarray}
  \label{helicity_dissipation}
  \frac{dK}{dt} & = & -2\eta \int_V j\cdot B\,\,dV\,\,\,\approx\,\,-\frac{2\eta}{\mu_0}B^2\frac{L^3}{l},\\
  \nonumber &  &\\
  \label{energy_dissipation}
  \frac{dW}{dt} & = & -\eta \int_V j\cdot j\,\,dV\,\,\,\approx\,\,-\frac{\eta}{\mu_0^2}B^2\frac{L^3}{l^2},
\end{eqnarray}
where \textit{j}\,=\,$\nabla\times B/\mu_0$ is the current density, \textit{l} is the length scale of magnetic variation (i.e., current sheet thickness), \textit{L} is the global length scale and $\eta$ is the resistivity (Browning, \citeyear{Browning1988}). Using $K\,\approx\,B^2 L$ and $W\,\approx\,B^2/2\mu_0$, the ratio of the dissipation rates reduces to $l/L$. Hence, $d_{t}K/K\,\ll\,d_{t}W/W$ if $l\,\ll\,L$, which is expected to be well satisfied for reconnecting current sheets within the highly conductive corona, where \textit{global} resistive diffusion of helicity and energy are negligible. The relative sizes of the dissipation rates have been confirmed by MHD simulations, despite the coarseness of numerical grids (the difference between dissipation rates becomes more pronounced as the resistivity becomes smaller and falls below numerical precision). Browning et al. (\citeyear{Browning2008}) showed that during the relaxation of a marginally (kink) unstable loop \textit{$\delta$K/K}\,$\sim$\,10$^{-4}$ and \textit{$\delta$W/W}\,$\sim$\,10$^{-2}$. Detailed estimates of coronal helicity dissipation are given by Berger (\citeyear{Berger1984}); further justification for helicity-conserving relaxation is provided by laboratory experiments (Heidbrink and Dang, \citeyear{Heidbrink2000}; Taylor, \citeyear{Taylor1986})

Helicity conservation combined with the invariant nature of the relaxed $\alpha$-profile imply that helicity has simply become more evenly distributed within the loop. Thus, the relaxed $\alpha$ can be calculated, as can the amount of energy liberated from the magnetic field during relaxation. The latter quantity can be interpreted as an upper limit to the heating event energy, since complete relaxation may not be attained. Issues concerning plasma response are outside the scope of the model presented in this paper.

Repeated episodes of slow photospheric driving may trigger an ideal MHD instability. Ideal (not resistive) instabilities are required in order for the time scale of the instability to match observations of the highly conducting corona, where resistive time scales are very long. Browning and Van der Linden (\citeyear{Browning2003}) describe how a dynamic heating event is caused whenever the field exceeds the threshold for a linear kink instability in a cylinderical coronal loop. Extensive numerical simulations (Galsgaard and Nordlund, \citeyear{Galsgaard1997}; Velli, Lionello, and Einaudi, \citeyear{Velli1997}; Lionello et al., \citeyear{Lionello1998}; Baty, \citeyear{Baty2000}; Gerrard et al., \citeyear{Gerrard2001}) have demonstrated how energy release occurs during the nonlinear phase of the instability. Relaxation of the field towards a constant-$\alpha$ state has also been demonstrated (Browning et al., \citeyear{Browning2008}; Hood, Browning, and Van der Linden, \citeyear{Hood2009}) alongwith a measurement of the energy release that is in good agreement with relaxation theory.

In summary, the model presented here is based on the idea that coronal loops are moved by photospheric motions through a series of force-free equilibria that will eventually culminate in a heating event whenever the ideal instability threshold is crossed. Bareford, Browning and Van der Linden (\citeyear{Bareford2010}, hereafter BBV\citeyear{Bareford2010}) calculated the heating event distributions for an ensemble of loops that possessed net current. This was done using a simple cylinderical field model in which the current profile ($\alpha$(\textit{r})) of the stressed field is represented by a two parameter family. (Theoretically, only two states on the closed instability threshold for two-parameter loops could yield zero-net-current configurations.) The resulting distributions were compatible with the energies required for coronal heating. Following BBV\citeyear{Bareford2010}, we use a simple cylindrical field model, and represent the nonlinear force-free field using a three-parameter family of current profiles in which $\alpha$ is a piecewise constant. We now attempt to improve the realism of the model by including a mechanism that requires each loop to have zero net current. Thus, we consider the effects of twisting motions localised within the loop cross section (Hood, Browning, and Van der Linden, \citeyear{Hood2009}). Previously, the random nature of photospheric motions was represented by allowing different parts of the loop interior to vary independently. It is more reasonable however to assume some level of correlation, since it is likely that the whole of a loop footpoint will be subjected to the same convective eddy. Furthermore, we also consider the consequence of varying the aspect ratio of the loop (\textit{L}\,/\textit{R}). This paper will also check to see if there is a twist-based parameter that can be used as a simple diagnostic for loop instability, e.g., Hood and Priest (\citeyear{Hood1979}).

The composition of the loop model and the equations used to express the magnetic field are given in the next section. Section \ref{sec:linear_kink_instability_thresholds} describes how the loop instability threshold is calculated and presents the equations used to determine the energy release associated with each relaxation. This section also analyses the loop configurations that follow the instability threshold. In addition, the loop's path to instability is explained. The nanoflare energy distributions produced by the model are presented in Section \ref{sec:distribution_of_energies_and_coronal_heating_considerations}. Finally, in the last section, the results are summarised and discussed.

\section{Model}
\label{sec:Model}
A loop is considered to evolve through equilibria as it is driven by slow photospheric footpoint motions. An idealised model of a straight cylindrical loop is used with the photosphere represented by two planes at \textit{z}\,=\,0,\,\textit{L}; however, the essential physics should apply to more complex geometries. The stressed field is line-tied with (in general) a non-uniform $\alpha$(\textit{r}) (where $\alpha\,=\,\mu_{0}\vec{j}_{\parallel}$/$\vec{B}$). This is represented by a three-parameter family of piecewise-constant-$\alpha$ profiles. 
\begin{figure}[h!]
  \center
  \includegraphics[scale=0.32]{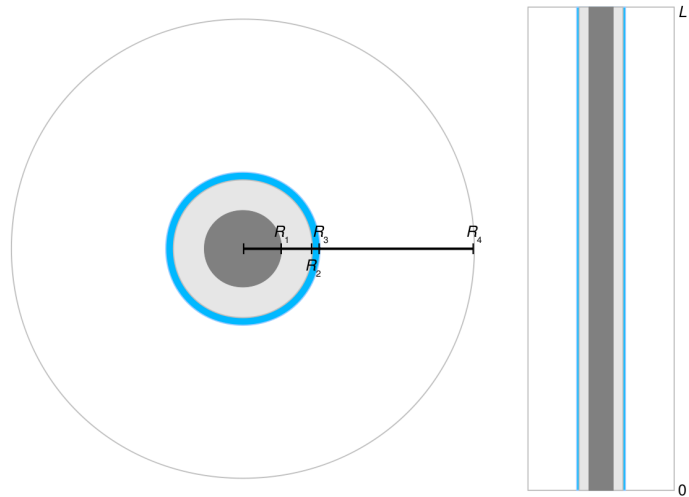}
  \caption{\small{Schematic of a straightened coronal loop in the \textit{r}-$\theta$ plane (left) and in the \textit{r}-\textit{z} plane (right). The loop, comprises a core (dark grey), an outer layer (light grey) and a current neutralisation layer (blue); the whole loop is embedded in a potential envelope (white). The core radius is half the loop radius and 1/6 the envelope radius (\textit{R}$_1$:\textit{R}$_2$:\textit{R}$_3$:\textit{R}$_4$ = 0.5:0.9:1:3). The loop aspect ratio (\textit{L}\,/\textit{R}$_3$) in this figure is 20.}}
  \label{schematic}
\end{figure}
The model discussed here (Figure \ref{schematic}) is an improved version of the one used by BBV\citeyear{Bareford2010}, which allowed loops to have net current (i.e., an azimuthal field was present in the potential envelope). A net current contradicts the idea that the magnetic field is twisted by convective motions local to the loop footpoints. If the twisting motions are confined to some localised region, the field surrounding the loop should be purely axial. The currents generated by the twisting of the fields within the loop should close locally such that the loop carries zero net current. Hence, a current neutralisation layer is introduced here, defined such that the azimuthal field ($B_{\theta}$) always falls to zero at the loop boundary (\textit{R}$_3$). Note that Hood, Browning and Van der Linden (\citeyear{Hood2009}) undertook 3D numerical simulations with initial fields taken as twisted states with zero net current. There are, however, only two such fields on the marginal instability curve of BBV\citeyear{Bareford2010}, for which the current due to $\alpha_2$ cancels that due to $\alpha_1$ (that is $\alpha_1$\,$\approx$\,$\mp$2.48, $\alpha_2$\,$\approx$\,$\pm$0.95). Here, we construct a whole family of current-neutralised fields by adding the extra layer.

Following on from BBV\citeyear{Bareford2010} and Browning and Van der Linden (\citeyear{Browning2003}), it is proposed that a relaxation event is triggered when the loop's field becomes linearly unstable. The energy released, due to fast magnetic reconnection during the nonlinear development of the instability, can then be calculated using relaxation theory.

The loop's radial $\alpha$-profile is approximated by a piecewise-constant function featuring three parameters. The ratio of current to magnetic field is $\alpha_1$ in the core, $\alpha_2$ in the outer layer, $\alpha_3$ in the neutralisation layer and zero in the potential envelope. The free parameters are $\alpha_1$ and $\alpha_2$, whereas $\alpha_3$ is dependent on the first two and is determined by the requirement of zero net current. Note that the magnetic field is continuous everywhere, whereas the current has discontinuities. Recent work indicates that a discontinuous current has little discernable effect on the dynamics when compared to similar but continuous $\alpha$-profiles (Hood, Browning, and Van der Linden, \citeyear{Hood2009}). Following the investigations of Browning et al. (\citeyear{Browning2008}), the outer surface of the potential envelope, representing the background corona, is placed at \textit{R}$_4$\,=\,3 (thrice the loop radius).

The fields are expressed in terms of the well-known Bessel function model, generalised to the concentric layer geometry (Melrose, Nicholls, and Broderick, \citeyear{Melrose1994}; Browning and Van der Linden, \citeyear{Browning2003}; Browning et al., \citeyear{Browning2008}; BBV\citeyear{Bareford2010}). Thus, the field equations for the four regions (core, outer layer, neutralisation layer and envelope) are as follows:
\begin{eqnarray}
  \label{field_equation_1}
  B_{1z} & = & B_{1}J_{0}(|\alpha_{1}|r)\\
  \label{field_equation_2}
  B_{1\theta} & = & \sigma_{1}B_{1}J_{1}(|\alpha_{1}|r) \mbox{\hspace{3.6cm}} 0 \leq r \leq R_{1}\\
  \label{field_equation_3}
  \nonumber &  &\\ 
  B_{2z} & = & B_{2}J_{0}(|\alpha_{2}|r) + C_{2}Y_{0}(|\alpha_{2}|r)\\
  \label{field_equation_4}
  B_{2\theta} & = & \sigma_{2}(B_{2}J_{1}(|\alpha_{2}|r) + C_{2}Y_{1}(|\alpha_{2}|r)) \mbox{\hspace{1cm}} R_{1} \leq r \leq R_{2}\\
  \nonumber &  &\\  
  \label{field_equation_5}
  B_{3z} & = & B_{3}J_{0}(|\alpha_{3}|r) + C_{3}Y_{0}(|\alpha_{3}|r)\\
  \label{field_equation_6}
  B_{3\theta} & = & \sigma_{3}(B_{3}J_{1}(|\alpha_{3}|r) + C_{3}Y_{1}(|\alpha_{3}|r)) \mbox{\hspace{1cm}} R_{2} \leq r \leq R_{3}\\
  \nonumber &  &\\  
  \label{field_equation_7}
  B_{4z} & = & B_{4}\\
  \label{field_equation_8}
  B_{4\theta} & = & 0 \mbox{\hspace{5.5cm}} R_{3} \leq r \leq R_{4}  
\end{eqnarray}
where $\sigma_i\,=\,\frac{\alpha_i}{|\alpha_i|}$ (\textit{i}\,=\,1,2,3) represent the sign of $\alpha_i$. The fields must be continuous at the inner radial boundaries, \textit{R}$_1$, \textit{R}$_2$ and \textit{R}$_3$. (We choose \textit{R}$_1$\,=\,0.5, \textit{R}$_2$\,=\,0.9 and \textit{R}$_3$\,=\,1, so that most of the loop is similar to previous work, BBV\citeyear{Bareford2010}, with a thin current neutralisation layer between \textit{R}$_2$ and \textit{R}$_3$) Therefore, the constants \textit{B}$_\textit{j}$ and \textit{C}$_\textit{j}$ (\textit{j} = 2,3,4) can be expressed like so:
\begin{eqnarray}
  \label{b2}
  B_2 & = & \frac{\pi |\alpha_2| B_1 R_1}{2}\Big(\sigma_{1,2}J_1(|\alpha_1|R_1)Y_0(|\alpha_2|R_1)-J_0(|\alpha_1|R_1)Y_1(|\alpha_2|R_1)\Big)\\
  \nonumber &  &\\
  \label{c2}
  C_2 & = & \frac{\pi |\alpha_2| B_1 R_1}{2}\Big(J_0(|\alpha_1|R_1)J_1(|\alpha_2|R_1) - \sigma_{1,2}J_1(|\alpha_1|R_1)J_0(|\alpha_2|R_1)\Big)\\
  \nonumber &  &\\  
  \label{b3}
  B_3 & = & \frac{\pi |\alpha_3| B_2 R_2}{2}\Big(\sigma_{2,3}F_1(|\alpha_2|R_2)Y_0(|\alpha_3|R_2) - F_0(|\alpha_2|R_2)Y_1(|\alpha_3|R_2)\Big)\\
  \nonumber &   &\\    
  \label{c3}
  C_3 & = & \frac{\pi |\alpha_3| B_2 R_2}{2}\Big(F_0(|\alpha_2|R_2)J_1(|\alpha_3|R_2) - \sigma_{2,3}F_1(|\alpha_2|R_2)J_0(|\alpha_3|R_2)\Big)\\    
  \nonumber &   &\\    
  \label{b4}
  B_4 & = & B_3 G_0(|\alpha_3|R_3)\\
  \nonumber &   &\\
  \label{c4}
  C_4 & = & 0
\end{eqnarray}
where
\begin{eqnarray}
  F_{0,1}(x) & = & J_{0,1}(x) + \frac{C_{2}}{B_{2}}Y_{0,1}(x)\\
  \nonumber & &\\
  G_{0,1}(x) & = & J_{0,1}(x) + \frac{C_{3}}{B_{3}}Y_{0,1}(x)
\end{eqnarray}
The value of $\alpha_3$ (the neutralisation layer current) is found, for a given ($\alpha_1$, $\alpha_2$), by numerical solution of \textit{B}$_{3\theta}$(\textit{R}$_3$)\,=\,0, ensuring that the net current is zero and that the azimuthal field vanishes outside the loop, see Equation \ref{field_equation_6}. Thus, the equilibrium parameter space remains 2D (i.e., it is determined by $\alpha_1$, $\alpha_2$) - the field profiles for a selection of loop configurations are given in Appendix B.

The magnetic flux through the loop and envelope is conserved:
\begin{eqnarray}
  \label{total_axial_flux}
  \nonumber \psi^{*} & = & \int_0^{R_4}2\pi r^* B_z^* \, dr^* = \frac{2\pi B_1 R_1}{|\alpha_1|}J_1(|\alpha_1|R_1)\\
  \nonumber &   & \mbox{\hspace{3cm}}+\frac{2\pi B_2 R_2}{|\alpha_2|}F_1(|\alpha_2|R_2)-\,2\pi R_1 J_1(|\alpha_1|R_1)\Bigg(\frac{\sigma_{1,2}}{|\alpha_2|}\Bigg)\\
  \nonumber &   & \mbox{\hspace{3cm}}+\,\frac{2\pi B_3}{|\alpha_3|}(R_3 G_1(|\alpha_3|R_3) - R_2 G_1(|\alpha_3|R_2))\\
  &   & \mbox{\hspace{3cm}}+\,\pi B_4\Big(R_4^2 - R_3^2\Big)    
\end{eqnarray}
where the asterisks denote dimensionless quantities. Hence, in the model, $\psi^{*}$ is normalised to 1 and \textit{B}$_1$ is determined (noting that, in Equations \ref{b2}-\ref{c4}, \textit{B}$_j$ and \textit{C}$_j$ are functions of $B_1$). The normalised coronal loop radius (\textit{R}$_3$\,=\,1) is itself used to normalise the loop length (e.g. \textit{L}\,=\,20\textit{R}$_3$), see Figure \ref{schematic}.

We assume that the loop evolves through a sequence of two-alpha fields (Equations \ref{field_equation_1}-\ref{field_equation_8}) as it is twisted by photospheric footpoint motions. It has been shown that the introduction of magnetic twist gives the coronal loop a circular cross section (Klimchuk, Antiochos, and Norton, \citeyear{Klimchuk2000}). The loop model presented here has the same cross-sectional shape, but the loop radius (\textit{R}$_3$) is held constant throughout the simulated photospheric driving. Purely azimuthal photospheric motions would cause a small expansion of the loop (Browning and Hood, \citeyear{Browning1989}) which we ignore; alternatively, small radial footpoint motions must be allowed in order to maintain constant loop radius. In any case, the sequence of loop equilibria explored by our model is clearly a small subset of the possible variation in field profiles that might arise from photospheric motions. As these random motions proceed, the loop continues to evolve through force-free equilibria until it becomes linearly unstable. We now discuss the calculation of the instability onset.

\section{Linear kink instability thresholds}
\label{sec:linear_kink_instability_thresholds}
A coronal loop's instability is constrained by the line-tying of the photospheric footpoints (Hood, \citeyear{Hood1992}). Hence, all perturbations are required to vanish at the loop ends (\textit{z}\,=\,0,\,\textit{L}). Any linear perturbation can be decomposed as a sum, $\sum\limits_{m=0}^{\infty} \tilde{f}(r,z)e^{i m\theta}e^{\gamma t}$. We need only consider the \textit{m}\,=\,1 term however, since this azimuthal mode has been found to be the least stable (Van der Linden and Hood, \citeyear{VanderLinden1999}) and is the dominant instability. The effect of such perturbations on the coronal loop are represented by the standard set of linearised ideal MHD equations. When the growth rate of a perturbation transitions from a negative value to a positive one, the loop has reached the threshold of an ideal linear instability. The instability threshold is a curve in 2-dimensional $\alpha$-space ($\alpha_1$, $\alpha_2$). The properties of the loop (e.g., $\alpha_1$, $\alpha_2$ and $\alpha_3$) at these threshold points can be found by substituting the perturbation function into the linearised MHD equations, leading to an eigenvalue equation for the growth rates (Priest, \citeyear{Priest1987}). The growth rates and eigenfunctions of the most unstable modes are found numerically, for line-tied fields, with the CILTS code, described in Browning and Van der Linden (\citeyear{Browning2003}) and Browning et al. (\citeyear{Browning2008}). CILTS can be configured such that one of the loop's free $\alpha$ parameters is fixed whilst the other is incremented. The code terminates as soon as the real part of the eigenfunction falls below zero, i.e., the loop is no longer unstable to kink perturbations.

\begin{figure}[h!]  
  \center
  \includegraphics[scale=0.65]{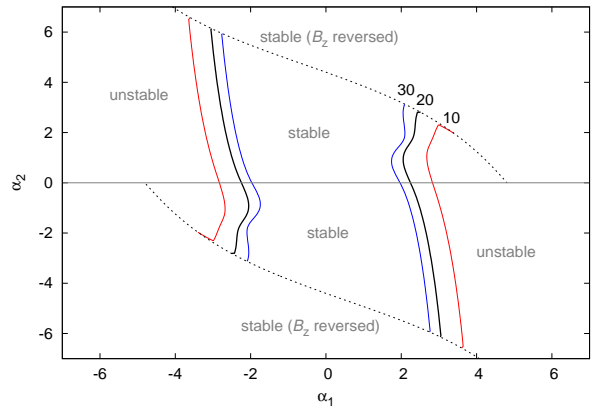}
  \caption{\small{The instability thresholds for \textit{L}\,/\textit{R}$_3$\,=\,10 (red), 20 (black) and 30 (blue). A closed stability region is formed by the \textit{B}$_z$ reversal lines (dashed).}}
  \vspace{-10pt}
  \label{its}
\end{figure}

Figure \ref{its} shows the instability threshold curves mapped by the CILTS code for three values of loop aspect ratio (\textit{L}\,/\textit{R}$_3$\,=\,10, 20, 30). The longer the loop the smaller the $\alpha$-value required for instability, since, if the radius is held constant, longer loops are less affected by line-tying stabilisations (Priest and Hood, \citeyear{Priest1979}). The addition of a current neutralisation layer prevents the threshold curves from closing near the $\alpha$-space origin; this is unlike the net current case for which the threshold is a closed curve (BBV\citeyear{Bareford2010}, Figure 2). The open shape is indeed similar to the instability curve for loops with a conducting wall at \textit{R}$_3$ (Browning and Van der Linden, \citeyear{Browning2003}). This is because the eigenfunctions almost vanish at \textit{R}$_3$, see Figures \ref{eigenfunctions1}-\ref{eigenfunctions3}. Also, if $\alpha_1$ is small, $\alpha_3$ will be opposite in sign to $\alpha_2$, and the outer layer is stabilised by the neutralisation layer. However, the loop configurations become unrealistic as we increase the magnitude of $\alpha_2$ and the axial field reverses. Positions outside the \textit{B}$_z$ reversal lines (Figure \ref{its}) represent loops that have axial fields of mixed polarity. These configurations cannot represent states attained by the twisting of a unipolar loop.

\begin{figure}[h!]  
  \includegraphics[scale=0.45]{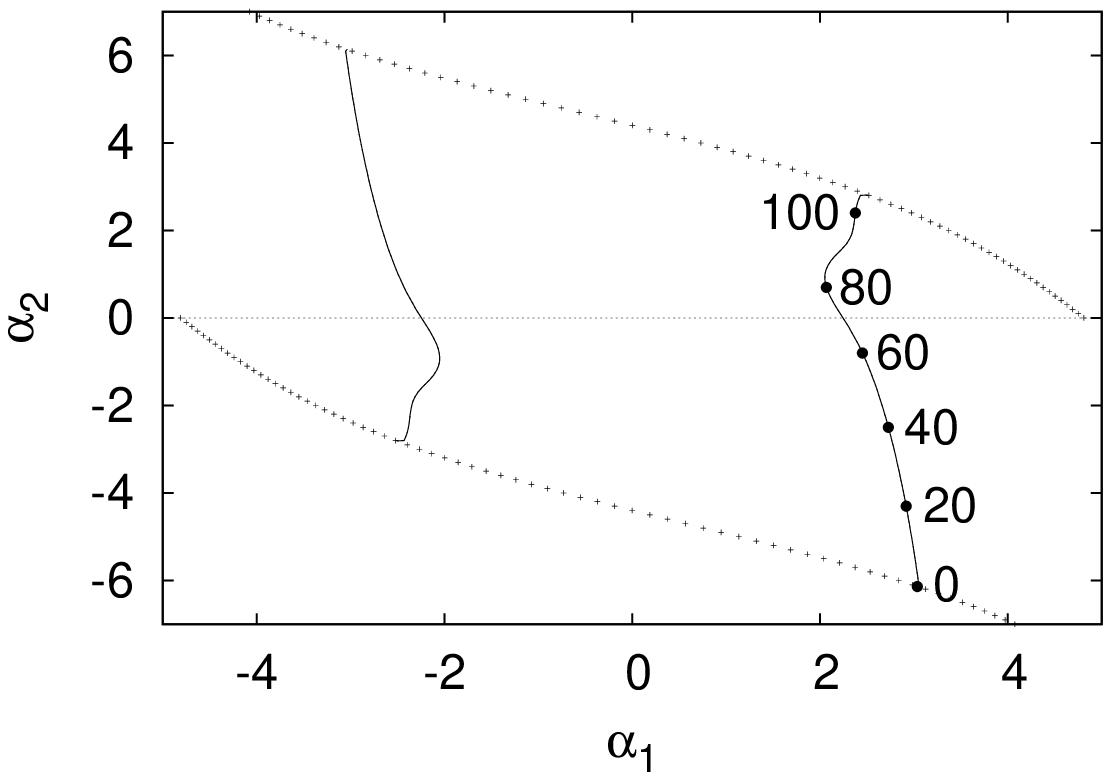}
  \includegraphics[scale=0.45]{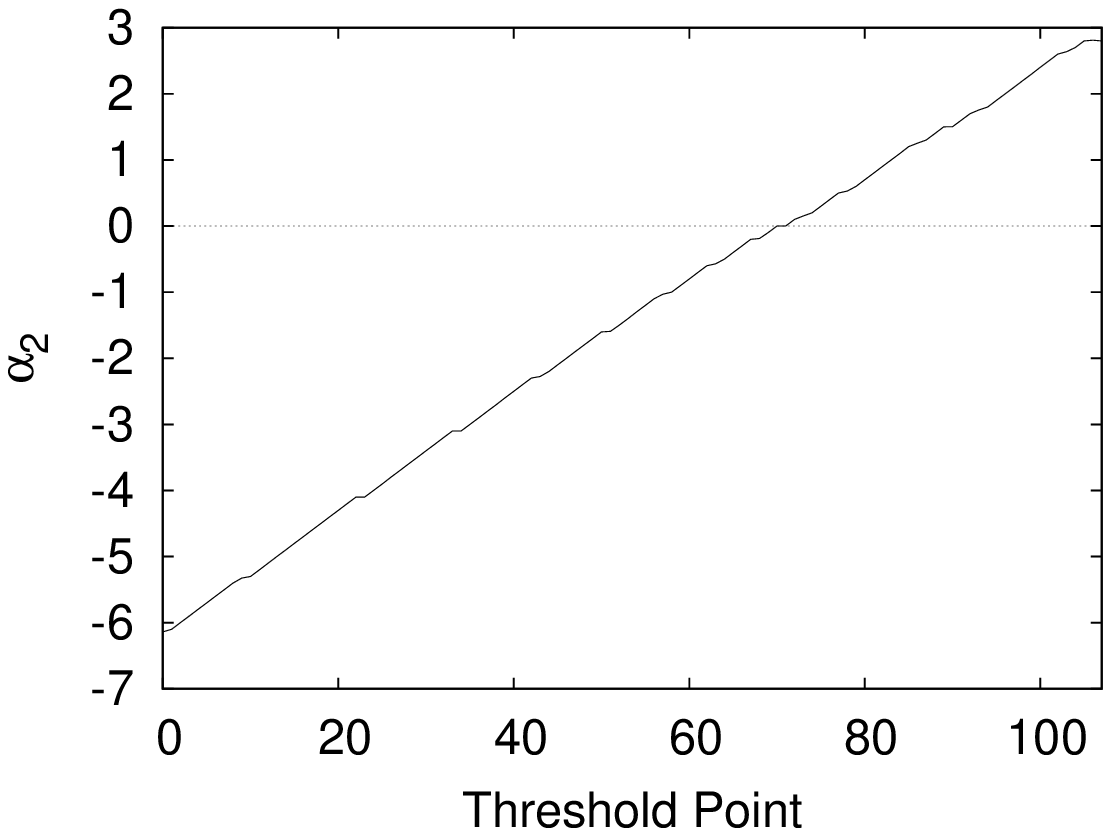}
  \caption{\small{On the left is the right half (i.e., where $\alpha_1$\,$\ge$\,0) of the instability threshold for a loop of aspect ratio 20. On the right is the variation in $\alpha_2$ along the 1D representation of the instability threshold. Notice that the tic marks along the Threshold Point axis correspond with the numbers that follow the threshold curves shown in the left plot.}}
  \label{it_a1p_ar20}
\end{figure}

\newpage
Before proceeding to calculate the energy release properties, it is first of interest to analyse the marginally unstable states. The threshold curves shown in Figure \ref{its} have symmetry: a rotation of $\pi$ radians leaves the thresholds unchanged. Thus, it is sufficient to show how various properties (e.g., magnetic twist and energy release) vary along the parts of the threshold where $\alpha_1$\,$\ge$\,0 (Figure \ref{it_a1p_ar20}). For ease of plotting we can convert the threshold curves to a one dimensional form: the filled circles and associated numbers of Figure \ref{it_a1p_ar20} (left) represent the tic marks and labels for the 1D threshold point axis, see Figures \ref{it_a1p_ar20_avtw_0_r3}-\ref{it_a1p_ar20_ax_wr}. Such figures will always be calculated using a loop of aspect ratio 20.

\begin{figure}[h!]
  \center  
  \includegraphics[scale=0.30]{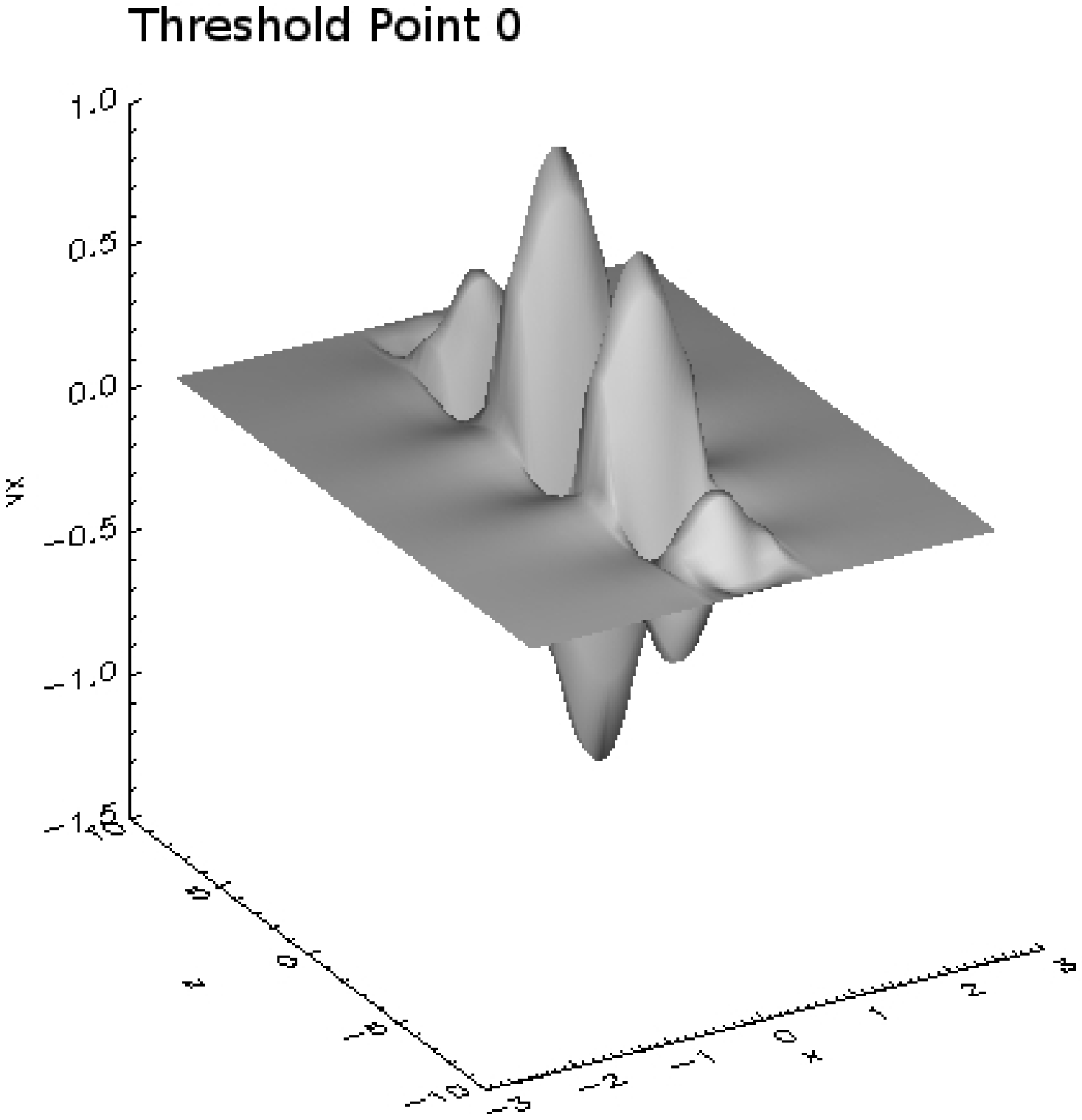}
  \includegraphics[scale=0.30]{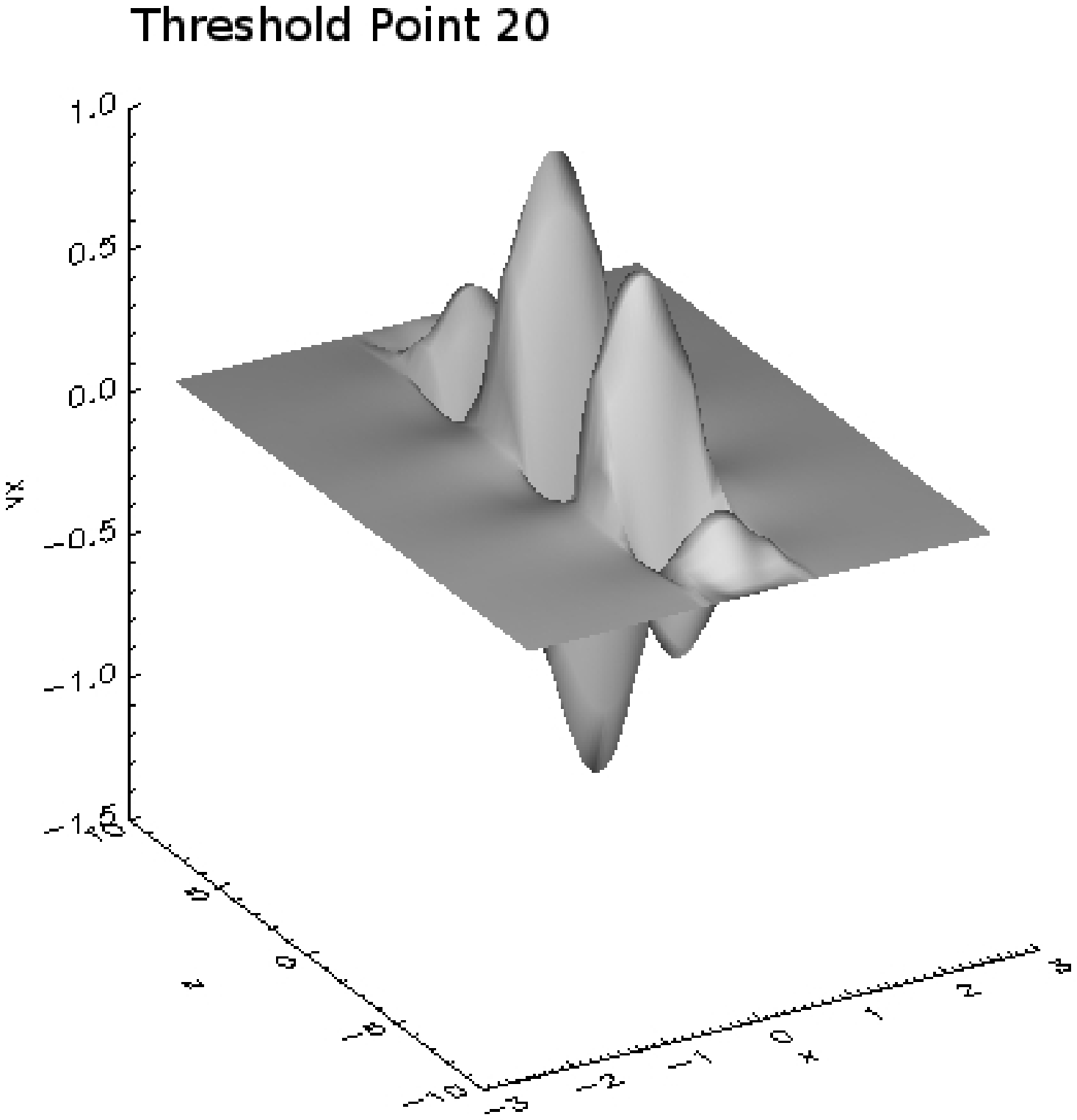}
  \caption{\small{The linear eigenfunctions, \textit{V}$_x$(\textit{x},\,\textit{y}=0,\,\textit{z}), for $\alpha$-space points 0 (left) and 20 (right). The $\alpha$ coordinates associated with each eigenfunction are on the unstable side of the threshold point number (Figure \ref{it_a1p_ar20}). Cartesian coordinates are used, hence, the x-axis is equivalent to the radial axis.}}
  \label{eigenfunctions1}
\end{figure}

\begin{figure}[h!]
  \center  
  \includegraphics[scale=0.30]{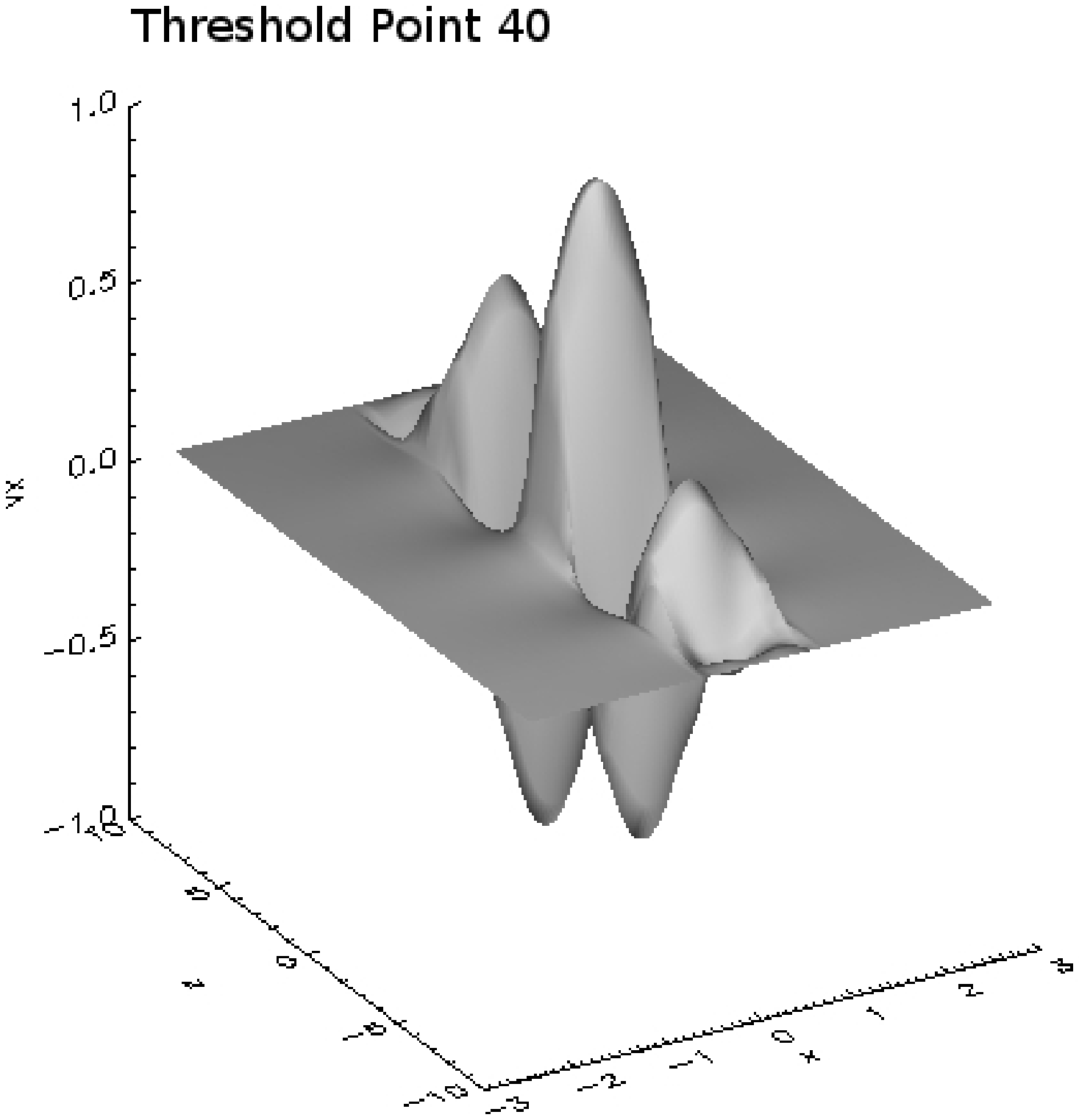}
  \includegraphics[scale=0.30]{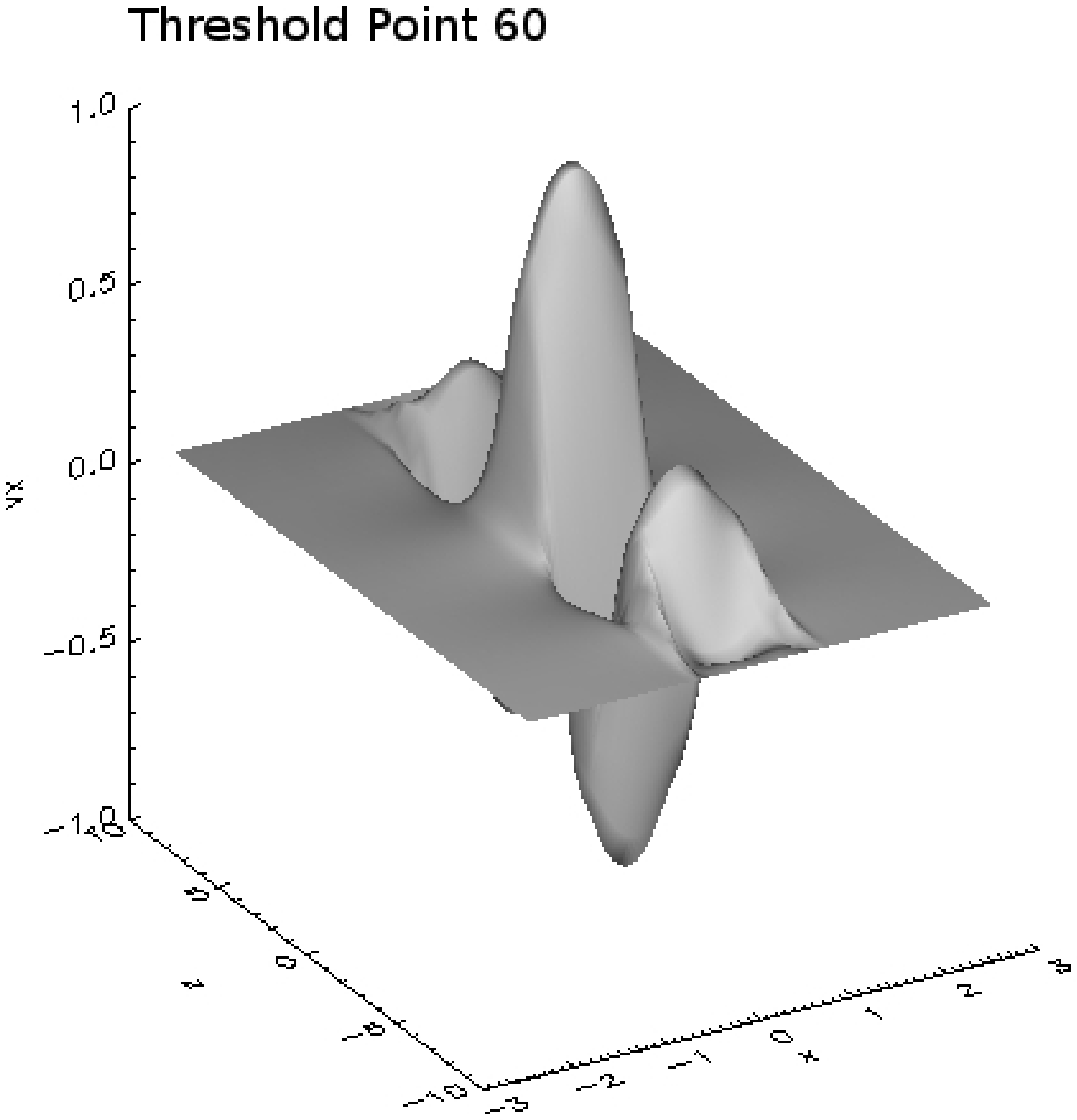}
  \caption{\small{The linear eigenfunctions, \textit{V}$_x$(\textit{x},\,\textit{y}=0,\,\textit{z}), for $\alpha$-space points 40 (left) and 60 (right).}}
  \label{eigenfunctions2}
\end{figure}

\newpage

Firstly, we plot the linear eigenfunctions for a selection of marginally unstable $\alpha$-space points that follow the instability threshold. The location of these points can be determined from the threshold point number given at the top of each plot in Figures \ref{eigenfunctions1}-\ref{eigenfunctions3}. The eigenfunctions of the marginally unstable modes are stongly radially confined; that is, there is almost no disturbance beyond the loop radius (\textit{R}$_3$\,=\,1). This contrasts sharply with the situation for loops with net current (Browning et al., \citeyear{Browning2008}; BBV\citeyear{Bareford2010}), in which the eigenfunction generally extends well into the potential envelope.

\begin{figure}[h!]
  \center  
  \includegraphics[scale=0.30]{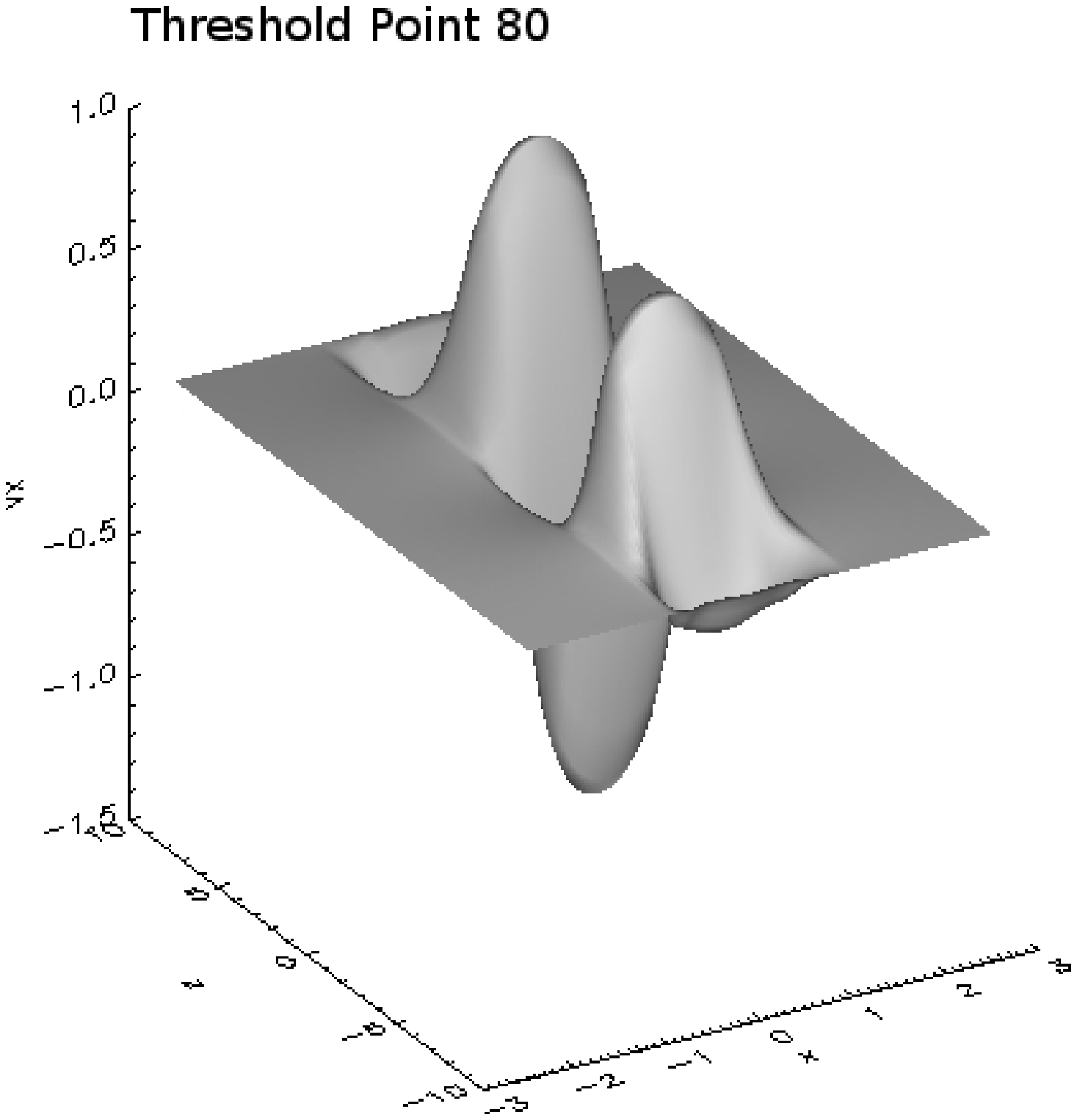}
  \includegraphics[scale=0.30]{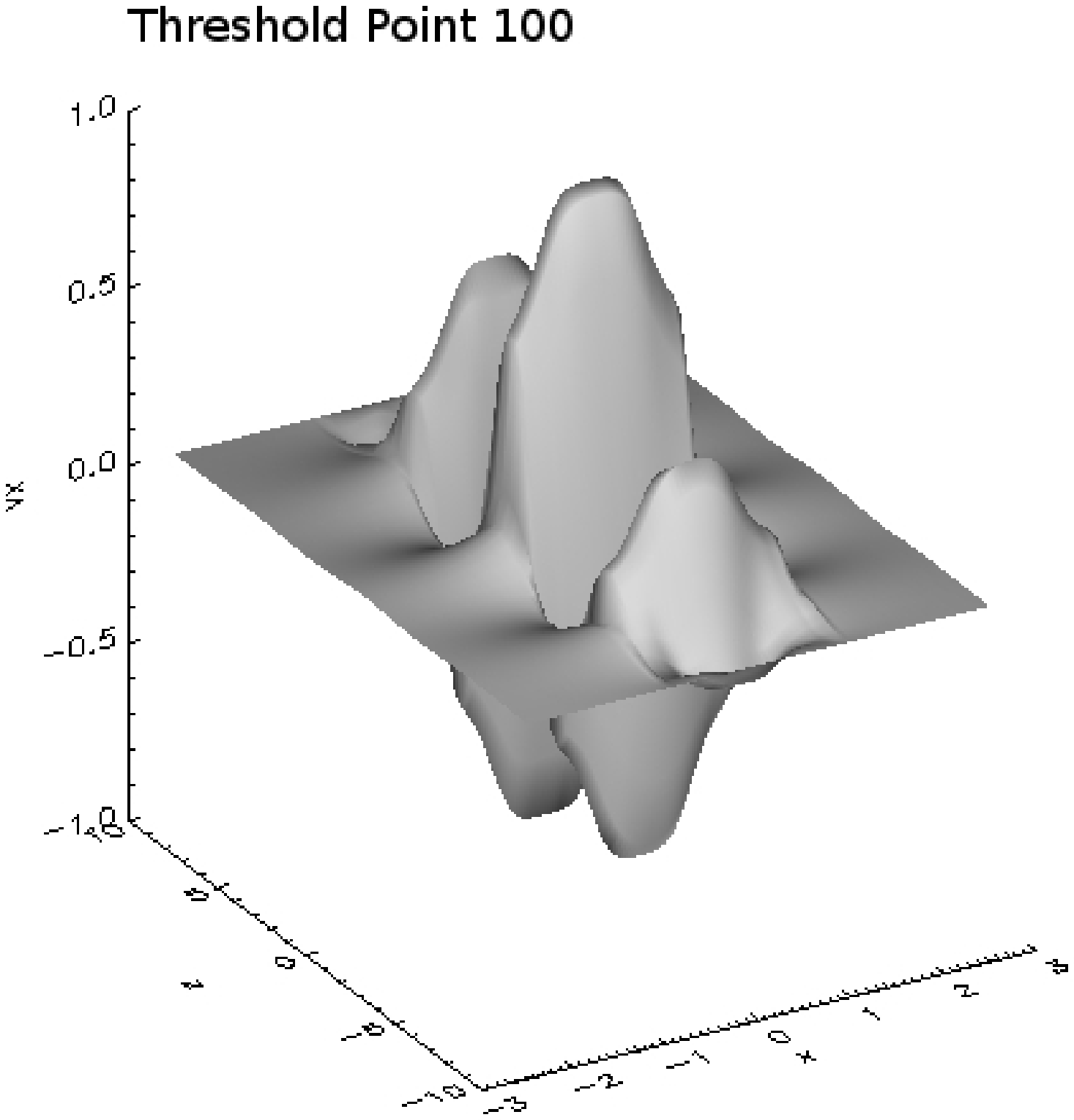}
  \caption{\small{The linear eigenfunctions, \textit{V}$_x$(\textit{x},\,\textit{y}=0,\,\textit{z}), for $\alpha$-space points 80 (left) and 100 (right).}}
  \label{eigenfunctions3}
\end{figure}

\subsection{Instability threshold and critical twist}
\label{sec:instability_threshold_and_critical_twist}
Following BBV\citeyear{Bareford2010}, we look for a single twist-related parameter that takes on a critical value whenever the loop reaches the threshold (Appendix B shows the twist profiles for a selection of loop configurations, stable and unstable). It has often been postulated that instability can be identified with a single \textit{critical twist} value irrespective of the detailed field profiles. The average twist can be calculated in several ways;
\begin{eqnarray}
  \label{average_velli_twist}
  {\langle\tilde{\varphi}\rangle}_0^{R_3} & = & \frac{\int_0^{R_3} LB_{\theta}(r)\,dr}{\int_0^{R_3} rB_z(r)\,dr}\,\,,\\
  \nonumber &   &\\
  \label{average_baty_twist}
  {\langle\hat{\varphi}\rangle}_0^{R_3} & = & \frac{1}{R_3}\int_0^{R_3} \frac{LB_{\theta}(r)}{rB_z(r)}\,dr\,\,,\\
  \nonumber &   &\\
  \label{average_wght_twist}
  {\langle\varphi\rangle}_0^{R_3} & = & \frac{1}{\pi R_3^{2}}\int_0^{R_3} 2\pi r \hspace{0.1cm} \frac{LB_{\theta}(r)}{rB_z(r)}\,dr\,\,.
\end{eqnarray}
Equation \ref{average_wght_twist} is the average twist weighted by area, while \ref{average_velli_twist} and \ref{average_baty_twist} have been used by Velli, Einaudi, and Hood (\citeyear{Velli1990}) and Baty (\citeyear{Baty2001}). Note, Equation \ref{average_velli_twist} can be calculated analytically, see Appendix A. ${\langle\varphi\rangle}_0^{R_3}$ denotes the average twist, weighted by area, over the core, outer layer and current neutralisation layer. The tilde ($\sim$) and hat ($\wedge$) symbols are used to indicate the other equations.
\begin{figure}[h!]
  \vspace{-5pt} 
  \center
  \includegraphics[scale=0.5]{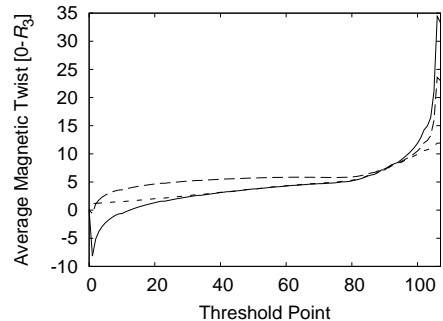}  
  \caption{\small{The variation in the loop average twist along the 1D representation of the instability threshold (\textit{L}\,/\textit{R}$_3$\,=\,20). The solid lines were calculated from Equation \ref{average_wght_twist}; the long dashed from Equation \ref{average_baty_twist} and the short dashed from Equation \ref{average_velli_twist}. The twist values are plotted in units of $\pi$.}}
  \label{it_a1p_ar20_avtw_0_r3}  
\end{figure}

None of the twist averages (Figure \ref{it_a1p_ar20_avtw_0_r3}) is invariant around the \textit{whole} threshold curve, although ${\langle\hat{\varphi}\rangle}_0^{R_3}$\,$\approx$\,5$\pi$ (Equation \ref{average_baty_twist}) for the majority of threshold points. This value is approximately in line with the oft-quoted result of 2.49$\pi$, the critical twist for a loop of aspect ratio 10 (Hood and Priest, \citeyear{Hood1981}). Each threshold point has a radial twist profile; these profiles feature reversed twist until around point 60, where the profile becomes single signed. After this point, the three average-twist plots converge to values between 5$\pi$ and 10$\pi$. At higher threshold points, the plots diverge and for Equation \ref{average_baty_twist} and \ref{average_wght_twist} the averages increase sharply.

Finally, we consider the proposal of Malanushenko et al. (\citeyear{Malanushenko2009}), that a critical value of normalised helicity (equivalent, in our terms, to the normalised loop helicity, $\textit{K}/\psi^2$, over the range 0\,-\,\textit{R}$_3$) indicates instability onset.  
\begin{figure}[h!]
  \center
  \includegraphics[scale=0.5]{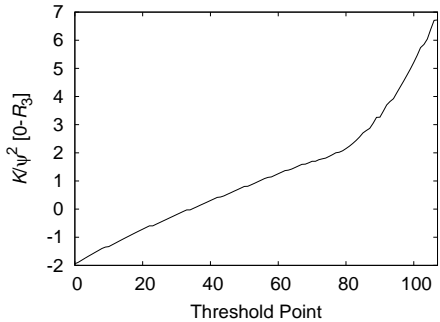}
  \caption{\small{The variation in \textit{K}/$\psi^2$ (over the range 0\,-\,\textit{R}$_3$) along the 1D representation of the instability threshold (\textit{L}\,/\textit{R}$_3$\,=\,20).}}
  \label{it_a1p_ar20_kopsi2_0_r3}  
\end{figure}
Figure \ref{it_a1p_ar20_kopsi2_0_r3} shows that the normalised helicity is certainly not the same for every threshold point, even if $\alpha_1$ and $\alpha_2$ have the same sign.

\subsection{Path to instability}
\label{sec:path_to_instability}
BBV\citeyear{Bareford2010} used a random walk process to simulate a loop being twisted by turbulent photospheric motions. In other words, a loop performed a sequence of fixed-length steps of random direction within $\alpha$-space until the instability threshold was crossed. We will follow this process for zero-net-current loops too, however, we will also employ spatially correlated random walks. This is to allow the correlation between the inner and outer parts of the loop to be varied. In particular, it is more likely that the twisting will be fairly uniform across the loop (i.e., the change in $\alpha_1$ is similar to the change in $\alpha_2$).

When a loop begins its random walk (i.e., when it emerges from beneath the photosphere) it is assigned a random starting position within the stable region of $\alpha$-space equilibria (i.e., the loop may have some initial twist). It is more likely however, that the initial twist will be small and that the initial value of $\alpha_2$ is similar to (or correlated with) the initial $\alpha_1$ value. Furthermore, the change in $\alpha$-coordinates that occur whenever the loop steps through $\alpha$-space, in response to photospheric driving, should also be correlated. The initial $\alpha_1$ coordinate of the walk is chosen from a normal distribution centred on zero. A standard deviation is chosen such that the probability of the initial $\alpha_1$ value representing an unstable configuration is negligible. Similarly, the initial $\alpha_2$ coordinate is chosen such that the mean is the initial $\alpha_1$ coordinate.

The step values, $\delta\alpha_1$ and $\delta\alpha_2$, are determined by assuming a step size, $\lambda$, and $\delta\alpha_1\approx\delta\alpha_2$. Hence, $\delta\alpha_1$ is also chosen from a normal distribution, but this time the mean is $\frac{\lambda}{\sqrt{2}}$ and $\delta\alpha_2$ is chosen such that the mean is $\delta\alpha_1$. 
\begin{figure}[h!]
  \center
  \includegraphics[scale=0.6]{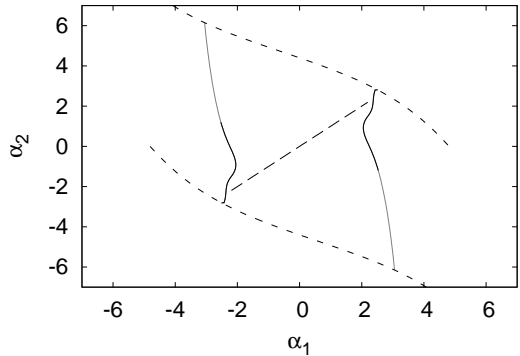}
  \caption{\small{The stability region for a loop of aspect ratio 20 is demarcated by instability thresholds (solid lines) and \textit{B}$_z$ reversal lines (short dashed lines). The loop configurations along the threshold have single-signed twist (black) or reversed twist (gray). The relaxation line (long dashed) comprises the points within the stability region where $\alpha_1$\,=\,$\alpha_2$.}}
  \label{corr_impact}  
\end{figure}
As the standard deviation of the normal distribution used to select $\delta\alpha_1$ and $\delta\alpha_2$ is decreased, the range of threshold crossings narrows. In other words, the walks follow the $\alpha_1$\,=\,$\alpha_2$ line more closely. A standard deviation of 0.1 will be used for the simulations presented in this paper, since this value restricts the threshold crossings to points where the twist is single-signed, see Figure \ref{corr_impact}. Correlated walks therefore are predisposed to maintaining the realism of loop configurations, since it is expected that in general photospheric motions do not create loops that have reversed twist.

Figure \ref{corr_impact} shows that a loop might cross a \textit{B}$_z$ reversal line before it reaches the instability threshold. If this happens, the loop is discarded and the simulation resumes with a new loop that has a stable $\alpha$-configuration. Once a loop reaches the instability threshold, it becomes linearly unstable. At this point, the field releases energy and transitions to a lower-energy state defined by Taylor relaxation: helicity is conserved and the $\alpha$-profile relaxes to a single value.

\subsection{Energy release calculation}
\label{sec:energy_release_calculation}
We allow each loop of an ensemble of 10$^5$ loops to undergo a single relaxation (BBV\citeyear{Bareford2010}). Initially, a loop starts from an assigned stable state. The field profile then undergoes a random walk (which may or may not be correlated) until it crosses the instability threshold; whereupon, the loop relaxes and the profile transitions to the relaxation line ($\alpha_1$\,=\,$\alpha_2$). The relaxation $\alpha$ ($\alpha_{x}$) will, of course, vary depending on where the threshold was crossed; $\alpha_{x}$ is found by helicity conservation (Taylor, \citeyear{Taylor1974}; Heyvaerts and Priest, \citeyear{Heyvaerts1984}; Taylor, \citeyear{Taylor1986}; Browning and Van der Linden, \citeyear{Browning2003}). In mathematical terms, we find the roots of the following equation:
\begin{eqnarray}
  \label{helicity_root}
  K(\alpha_x) - K(\alpha_{i1},\alpha_{i2}) & = & 0,
\end{eqnarray}
where $\alpha_{i1}$ and $\alpha_{i2}$ are the coordinates of the instability threshold crossing (conservation of axial flux is assured through the normalisation $\psi^{*}$\,=\,1). The helicity can be expressed as follows:
\begin{eqnarray}
  \label{helicity}
  K & = & 2L \int_{0}^{R_x}\frac{I(r)\psi (r)}{r} \, dr,  
\end{eqnarray}
where \textit{I}(r) is the axial current and \textit{L} is the loop length (Finn and Antonsen, \citeyear{Finn1985}). Axial flux is also conserved. The full expressions for helicity and energy are given in Appendix A. The energy difference between the unstable and relaxed states can be calculated thus:
\begin{eqnarray}
  \label{energy_release}
  \delta W & = & W(\alpha_{i1},\alpha_{i2}) - W(\alpha_{x}).
\end{eqnarray}
This is the relaxation energy: the energy that is liberated from the magnetic field during the event. How much of this energy is converted to heat depends on the plasma response; thus, the relaxation energy represents the upper limit of the energy available for plasma heating.

In BBV\citeyear{Bareford2010}, a loop with net current was relaxed such that the $\alpha$-profile became invariant over the range 0-\textit{R}$_4$. Hence, the relaxed state always represented a threefold radial expansion of the threshold state (i.e., from \textit{R}$_3$ to \textit{R}$_4$), the relaxation encompassed both the loop and the potential envelope. Numerical simulations (Browning et al., \citeyear{Browning2008}) indicate that this is a good model for loops with net current. However, for loops with zero net current, the instability is more radially confined and the reconnection activity is correspondingly localised; it does not extend to the outer boundary (Hood, Browning, and Van der Linden, \citeyear{Hood2009}; Bareford et al., \citeyear{Bareford2011}).

We therefore consider that the relaxation radius, \textit{R}$_x$, can be anywhere in the range \textit{R}$_3$(=\,1)\,$\leq$\,\textit{R}$_x$\,$\leq$\,\textit{R}$_4$(=\,3). If \textit{R}$_x$\,=\,\textit{R}$_4$, we have \textit{complete relaxation} as previously considered (Browning and Van der Linden, \citeyear{Browning2003}; Browning et al., \citeyear{Browning2008}; BBV\citeyear{Bareford2010}); otherwise relaxation is localised. $\alpha$ is constant between 0 and \textit{R}$_x$ and the fields in the remaining envelope (where $\alpha$\,=\,0 and \textit{R}$_x$\,$\leq$\,\textit{r}\,$\leq$\,\textit{R}$_4$) are fixed so that they do not change during relaxation; this maintains current neutralisation, albeit via an infinitely thin current-neutralising surface. Axial flux is conserved, such that $\psi$ (over 0-\textit{R}$_x$) of the threshold state is equal to $\psi$ (over 0-\textit{R}$_x$) of the relaxed state. \textit{K}/$\psi^2$ is conserved in an identical manner (in our previous work, conservation was always over 0-\textit{R}$_4$ and since the total axial flux was normalised to 1, conserving \textit{K}/$\psi^2$ was identical to conserving \textit{K}). Likewise, the energy release is the energy of the threshold state over 0-\textit{R}$_x$ minus the energy of the relaxed state over the same radial range. In fact, the energy of the remaining potential envelope is unchanged, so that the energy release could also be taken over the entire volume (0-\textit{R}$_4$); similarly, the envelope has zero helicity before and after relaxation.

\section{Distribution of energies and coronal heating considerations}
\label{sec:distribution_of_energies_and_coronal_heating_considerations}
We now proceed to the main task of our work, which is to calculate the distribution of magnitudes of the sequence of heating events generated by random photospheric driving. First, we show how various properties vary along the instability threshold.

\subsection{Helicity and energy}
\label{sec:helicity_and_energy}
\begin{figure}[h!]
  \center
  \includegraphics[scale=0.45]{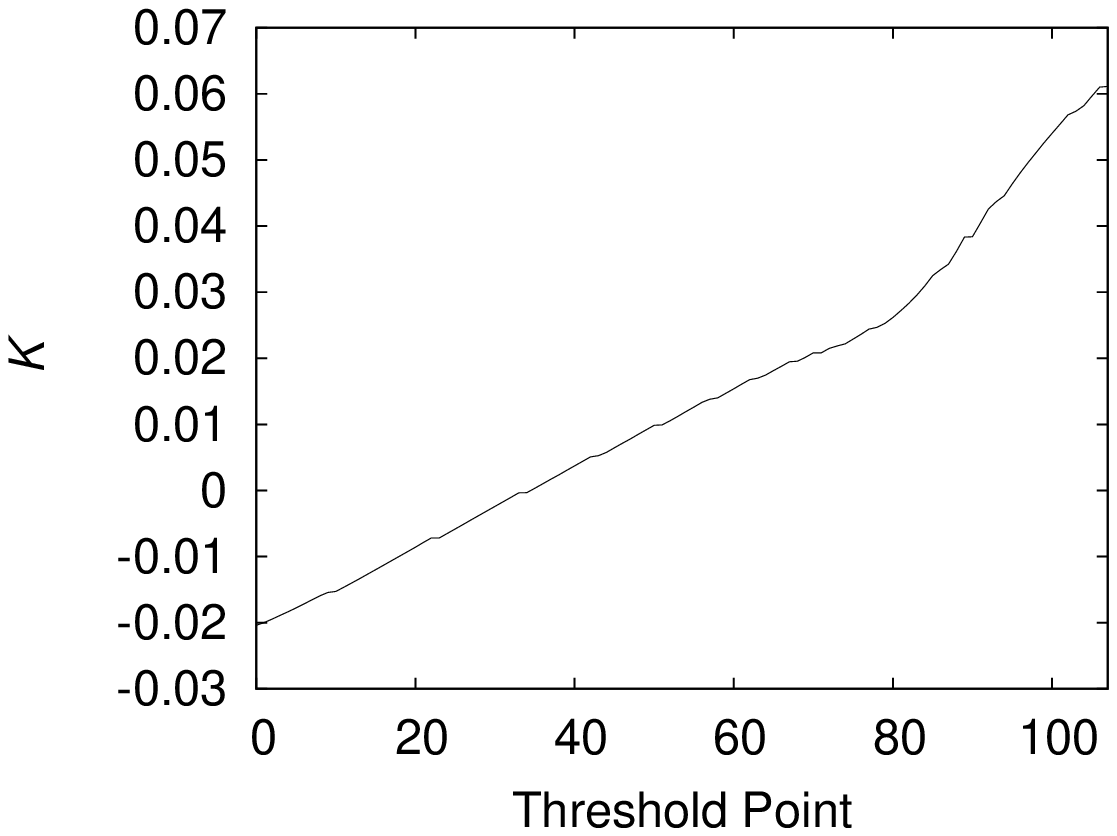}
  \includegraphics[scale=0.45]{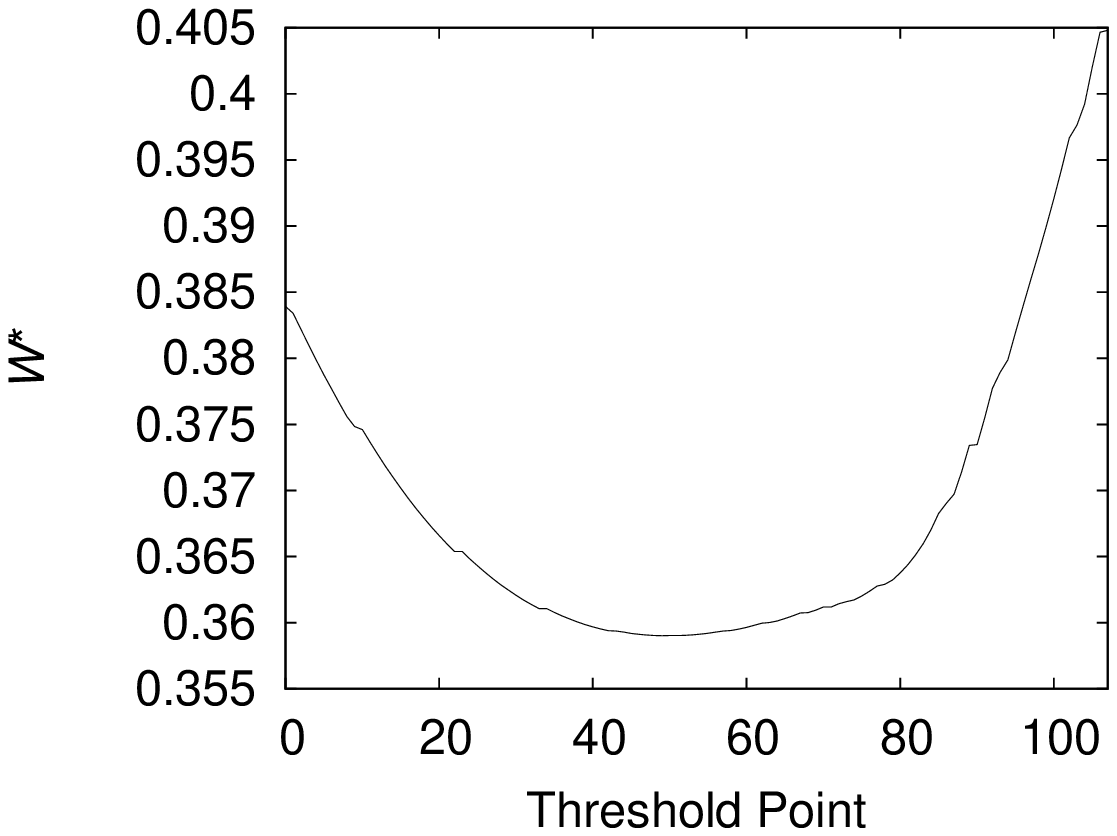}
  \caption{\small{Total helicity (left) and total (dimensionless) magnetic energy (right) along the 1D representation of the instability threshold (\textit{L}\,/\textit{R}$_3$\,=\,20).}}
  \label{it_a1p_ar20_k_w}  
\end{figure}

The left panel of Figure \ref{it_a1p_ar20_k_w} plots total helicities of the threshold states. A total helicity (or flux) is one calculated over the range 0\,-\,\textit{R}$_4$, i.e., the loop and envelope. 
None of the threshold states have sufficient helicity for the relaxed state to feature helical modes (Taylor, \citeyear{Taylor1986}) and so all relaxed states are cylinderically symmetric.
\begin{figure}[h!]
  \center
  \includegraphics[scale=0.45]{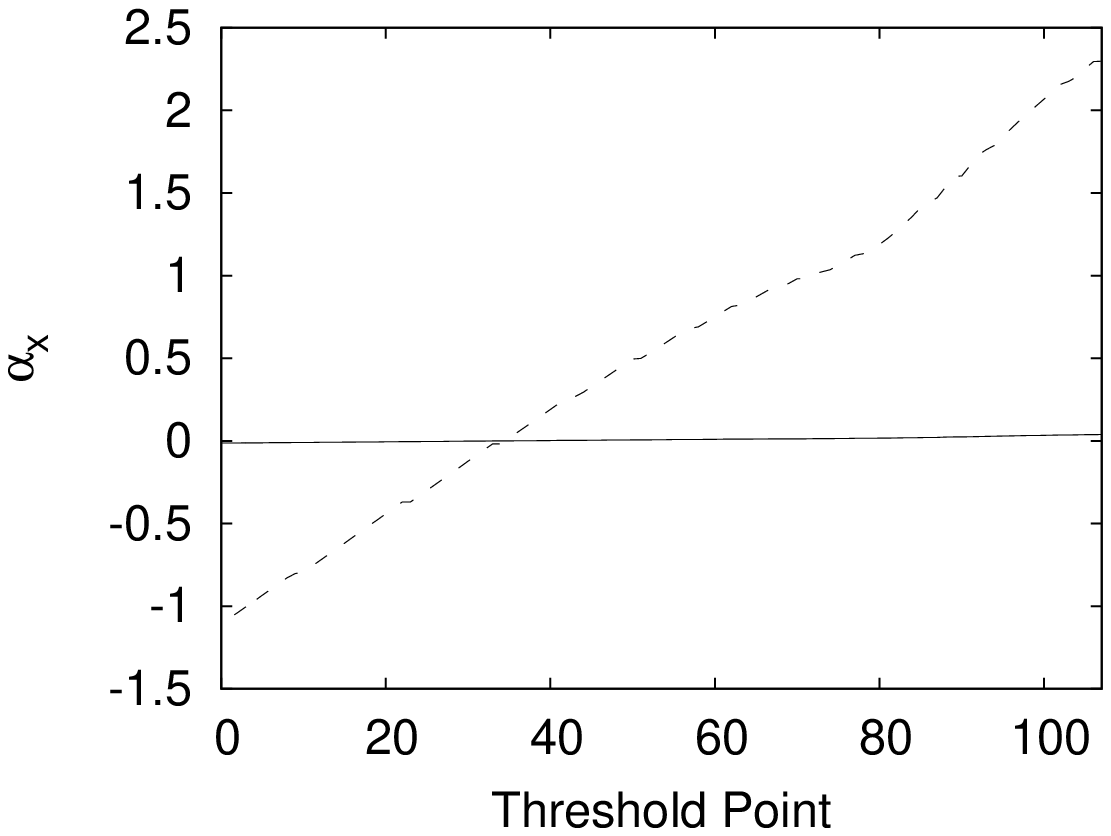}
  \includegraphics[scale=0.45]{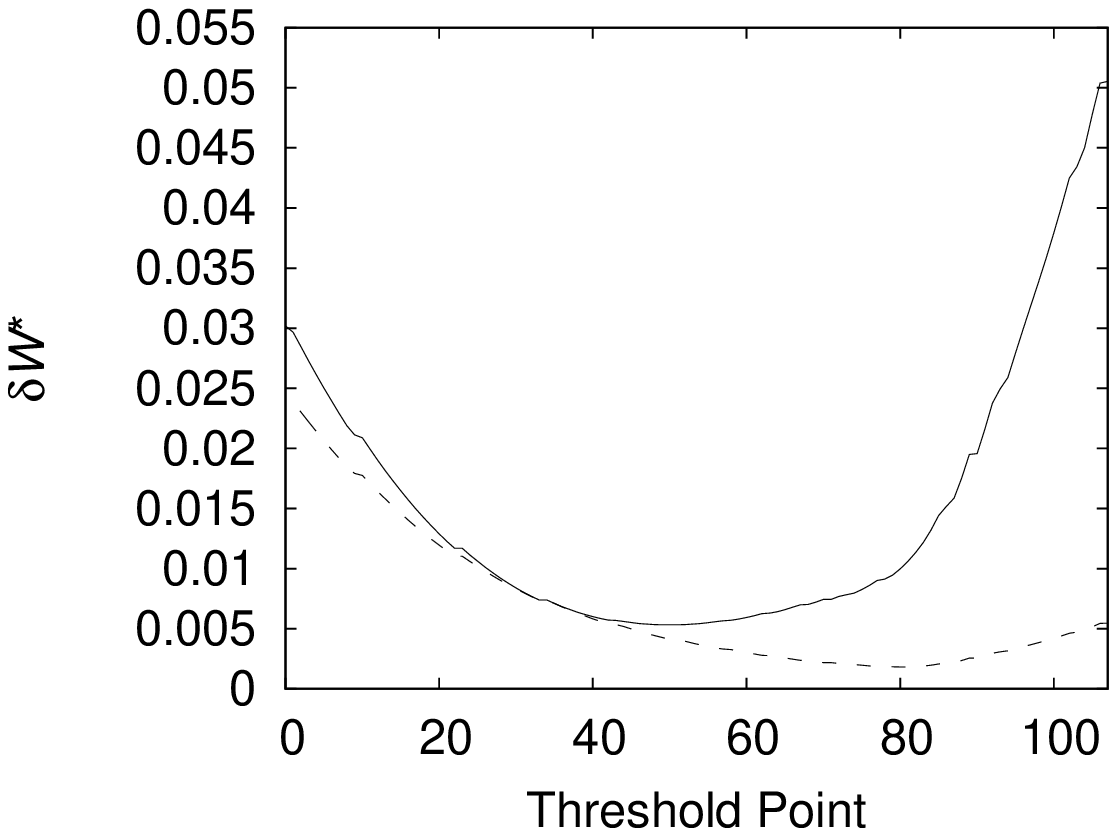}
  \caption{\small{$\alpha_x$ (left) and energy release (right) along the 1D representation of the instability threshold (\textit{L}\,/\textit{R}$_3$\,=\,20). These properties have been calculated for two relaxation radii \textit{R}$_x$\,=\,\textit{R}$_3$ (dashed) and \textit{R}$_x$\,=\,\textit{R}$_4$ (solid). When \textit{R}$_x$\,=\,\textit{R}$_4$, $\alpha_x$ is of \textit{O}(10$^{-2}$) and so the corresponding plot appears very close to the $\alpha_x$\,=\,0 line.}}
  \label{it_a1p_ar20_ax_wr}  
\end{figure}
The left panel of Figure \ref{it_a1p_ar20_ax_wr} confirms that each threshold state corresponds to a relaxed state. Both of the graphs in Figure \ref{it_a1p_ar20_ax_wr} feature two plots; the dashed line represents minimum relaxation (\textit{R}$_x$\,=\,\textit{R}$_3$) and the solid line represents full relaxation (\textit{R}$_x$\,=\,\textit{R}$_4$). The right panel shows that, in general, $\delta$\textit{W}$^*$ is affected by \textit{R}$_x$ (although, there is one part of the threshold where the energy release is insensitive to relaxation radius); hence, the energy distributions that appear in the next section are calculated for minimum and maximum relaxation radii.

The energies shown in Figure \ref{it_a1p_ar20_k_w} (right) and Figure \ref{it_a1p_ar20_ax_wr} (right) are given as dimensionless quantities; BBV\citeyear{Bareford2010} derived the following expression for calculating a dimensional energy,
\begin{eqnarray}
  \label{dim_energy_release}	
  \delta W & = & 81\frac{\pi^2}{\mu_0} R_c^3 B_c^2 \hspace{0.1cm} \delta W^{\hspace{0.02cm}*}\hspace{0.1cm},
\end{eqnarray}
where \textit{R}$_c$ is the loop radius in the corona and \textit{B}$_c$ is the mean axial field in the corona. Assuming typical values (\textit{R}$_c$\,=\,1 $\mathrm{Mm}$ and \textit{B}$_c$\,=\,0.01 $\mathrm{T}$), we obtain dimensional energy values of $6\,\times\,10^{22}\,\delta\textit{W}^{\,*}$ $\mathrm{J}$\,\,$\equiv$\,\,$6\,\times\,10^{29}\,\delta\textit{W}^{\,*}$ $\mathrm{erg}$. Thus, the top end of the $\delta$\textit{W}$^{\,*}$ scale ($\approx$\,0.05) for \textit{R}$_x$\,=\,3 is equivalent to $3\,\times\,10^{28}$ $\mathrm{erg}$. This is in the microflare range, but nanoflare energies will be obtained for weaker fields.

\subsection{Flare energy distribution}
\label{sec:flare_energy_distribution}
An expression for the energy flux is derived by considering the loops in the ensemble as spatially separated but flaring simultaneously. All the energy input from the photosphere is dissipated, in a long-term time-average over many events, since the instability threshold limits the accumulation of stresses within the coronal magnetic field.
The energy flux, \textit{F}, is thus;
\begin{eqnarray}
  \label{dim_energy_flux}	
  F & = & \frac{81}{2}\frac{\pi}{\mu_0} R_c B_c^2 \hspace{0.1cm} \frac{1}{N\tau} \hspace{0.1cm} \frac{1}{10^5} \sum_{i=1}^{10^5}\delta W^{\hspace{0.02cm}*},
\end{eqnarray}
where \textit{N} is the average number of steps taken to reach the threshold and $\tau$ is the time taken to complete each step in the random walk. Similar expressions exist in the literature based on random photospheric twisting (Sturrock and Uchida, \citeyear{Sturrock1981}; Berger, \citeyear{Berger1991}; Zirker and Cleveland, \citeyear{Zirker1993}; Abramenko, Pevstov, and Romano, \citeyear{Abramenko2006}).

In other words, $\tau$ is the time taken for $\alpha$ to change by $\lambda$\,/\textit{R}$_3$; we may estimate, a timescale for this process as follows. Based on axial values, $\lambda$ corresponds to a change in magnetic twist $\delta\phi$\,$\approx$\,(\textit{L}/2)($\lambda$\,/\textit{R}$_3$); taking \textit{L}\,/\textit{R}$_3$\,=\,20 and $\lambda$\,=\,1 gives $\delta\phi$\,$\approx$\,10. If this is caused by photospheric twisting motions of magnitude \textit{v}$_{\theta}$ for a time interval $\tau$, we find $\tau$\,$\approx$\,($\delta\phi$)\textit{R}$_f$/\textit{v}$_{\theta}$, where \textit{R}$_f$ is the footpoint radius (at the photosphere). With typical values of \textit{R}$_f$\,=\,200\,$\mathrm{km}$ and \textit{v}$_{\theta}$\,=\,1\,$\mathrm{km\,\,s^{-1}}$, we obtain $\tau$\,$\approx$\,2000 $\mathrm{s}$; note that this is consistent with the expected correlation time for photospheric motions; granule lifetimes are of the order 10$^3$ $\mathrm{s}$ (Zirker and Cleveland, \citeyear{Zirker1993}). Hence $\tau$ has a linear relationship with the loop length, $\tau$\,=\,100(\textit{L} [$\mathrm{Mm}$]). Applying the previously used values for \textit{B}$_c$ and \textit{R}$_c$, gives a dimensional flux of $(10^{8}/N\tau)\sum\delta W^{\hspace{0.02cm}*}$ $\mathrm{erg\,\,cm^{-2}\,\,s^{-1}}$. This result is applicable to Active Regions (a value for the Quiet Sun can be obtained by setting \textit{B}$_c$\,=\,0.001 $\mathrm{T}$; this simply lowers the multiplier (10$^{8}$) by 2 orders of magnitude.

The energy distributions given below are each derived from a loop ensemble, that is a collection of 10$^5$ loops flaring simultaneously. Since each loop relaxes only once we can sidestep the complications that come with allowing loops to survive many relaxations: for example, a loop may shrink or implode after flaring (Janse and Low, \citeyear{Janse2007}), and a different instability threshold would have to be applied should the aspect ratio be altered as a consequence. 

\subsubsection{Distribution of "nanoflares"}
\label{sec:distribution_of_nanoflares}

\begin{figure}[h!] 
  \center
  \includegraphics[scale=0.44]{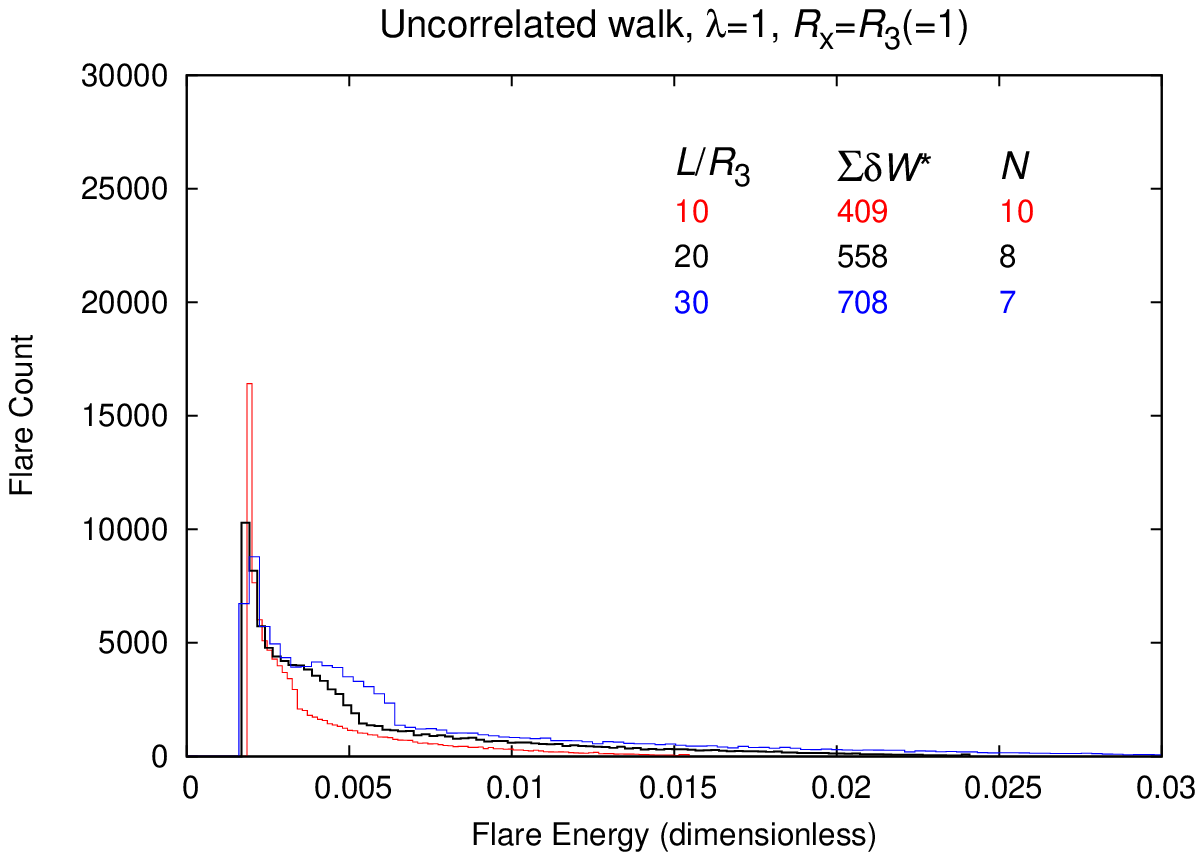}
  \includegraphics[scale=0.44]{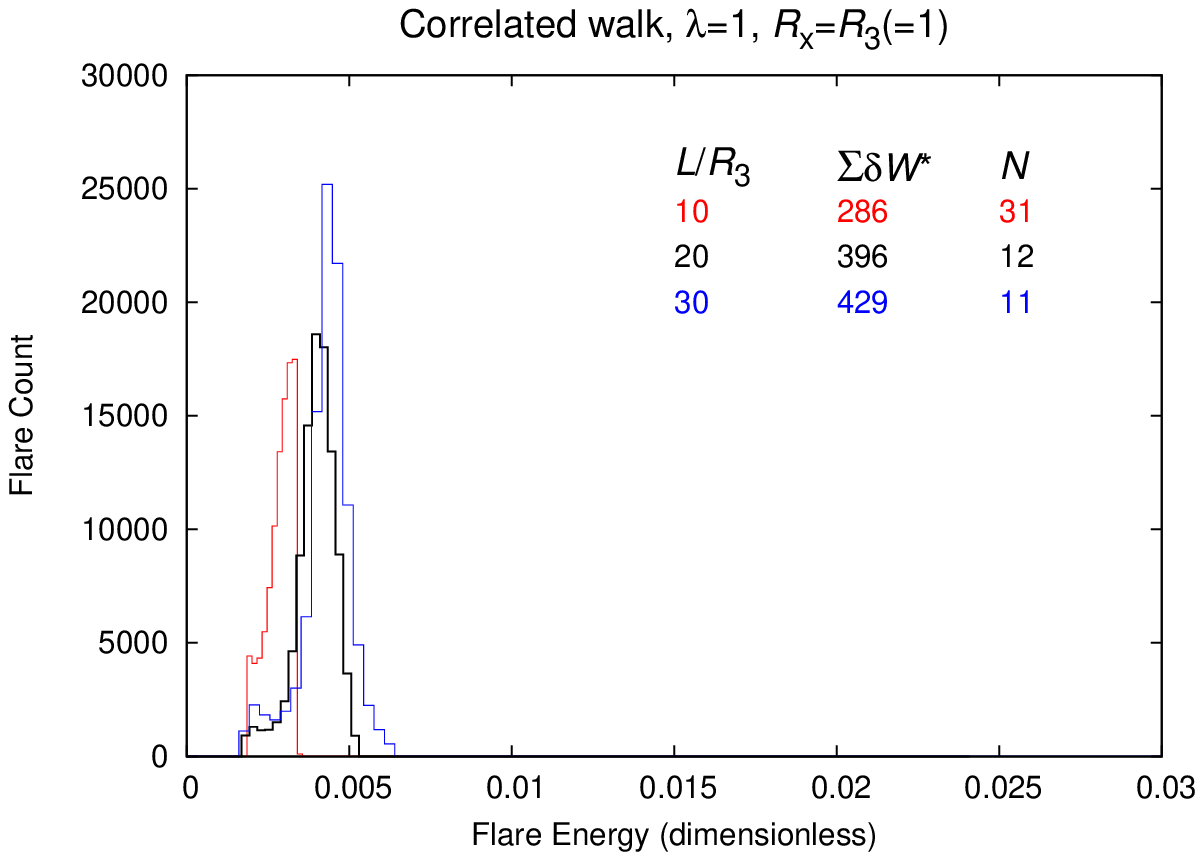}
  \includegraphics[scale=0.44]{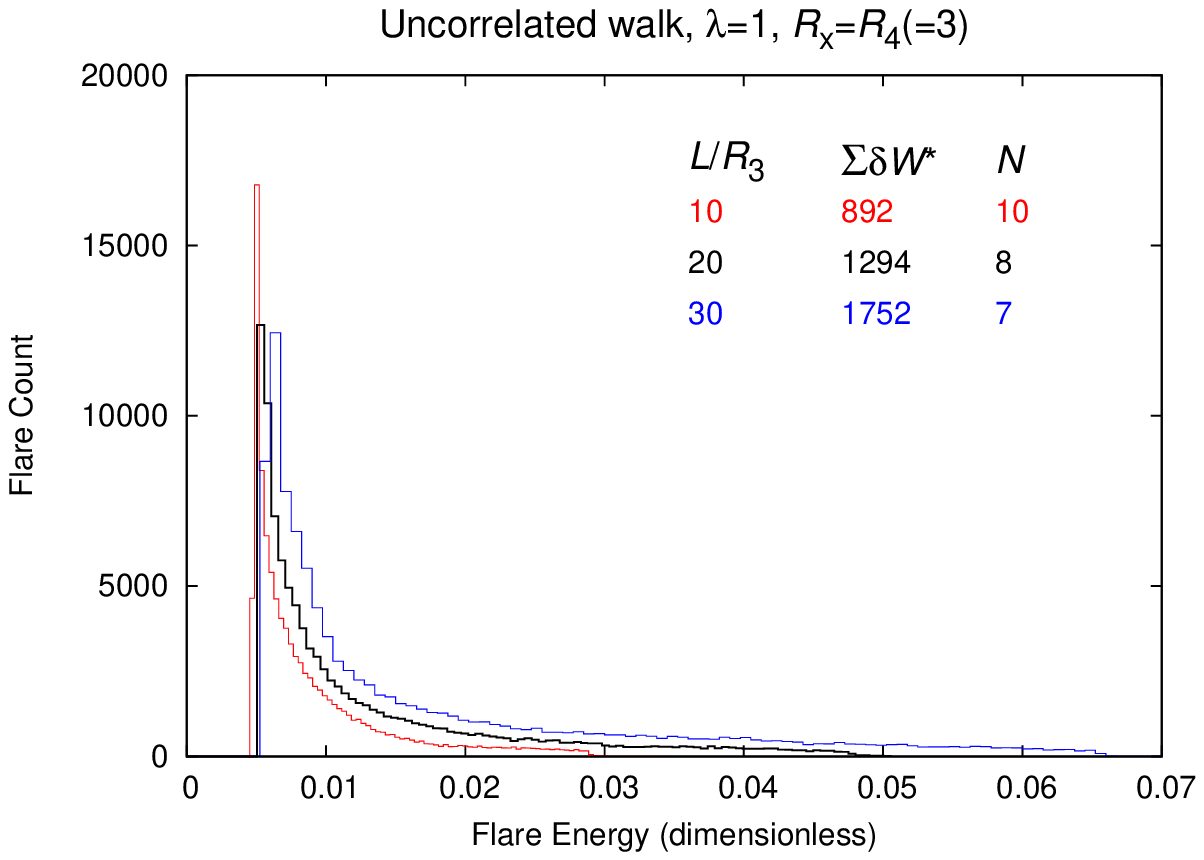}
  \includegraphics[scale=0.44]{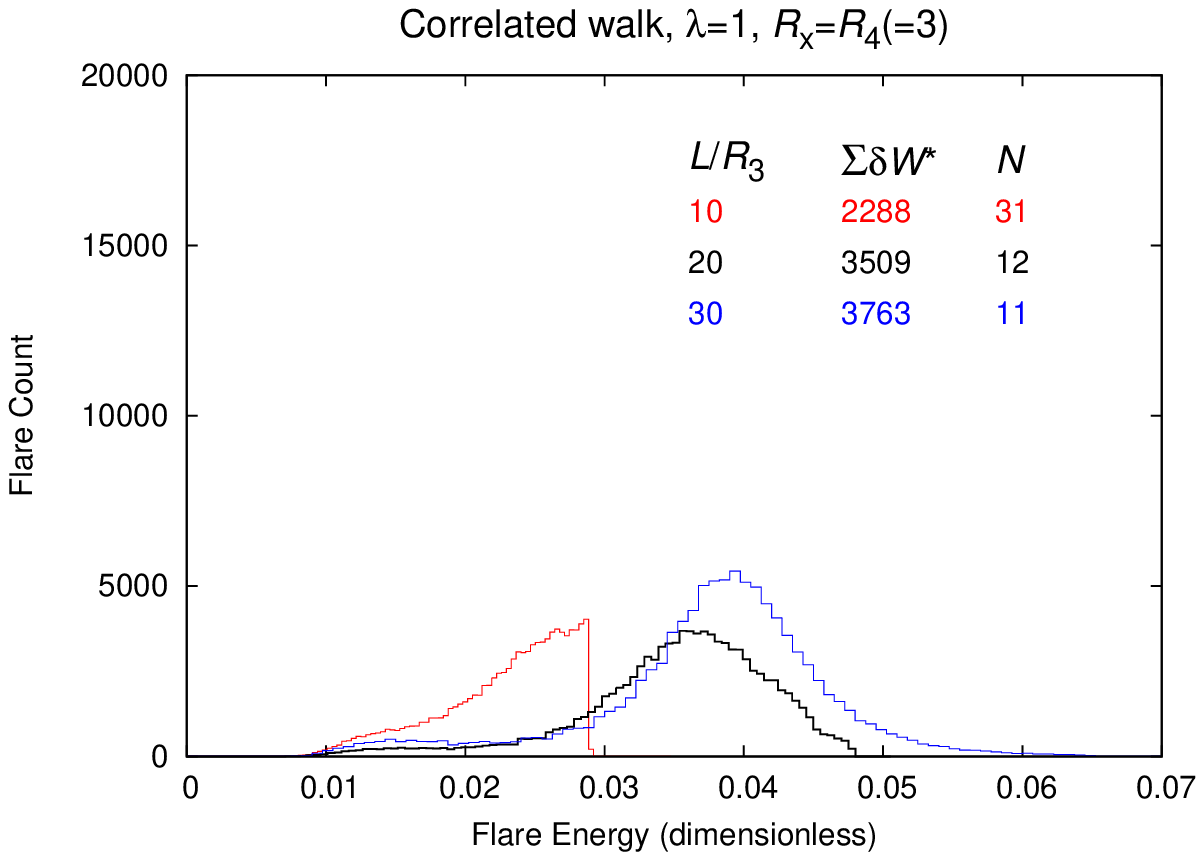}
  \caption{\small{Flare energy distributions for a 10$^5$ loop ensemble, with each loop undergoing one relaxation event. The relaxation radius (\textit{R}$_x$) associated with each event is \textit{R}$_3$ for the top two and \textit{R}$_4$ for the bottom two. The plots on the left correspond to uncorrelated random walks, those on the right to correlated driving. The distribution curves are colour-coded according to aspect ratio: red denotes \textit{L}\,/\textit{R}$_3$\,=\,10, black \textit{L}\,/\textit{R}$_3$\,=\,20 and blue \textit{L}\,/\textit{R}$_3$\,=\,30. In addition, two properties are displayed for each plot, $\sum\delta W^{\hspace{0.02cm}*}$, the total energy release (dimensionless) and \textit{N}, the average number of steps taken to reach the threshold.}}
  \label{wrpf}
\end{figure}

Examination of Figure \ref{wrpf} yields three key points. Firstly, the total energy released increases with aspect ratio, but the average step count, \textit{N}, decreases. This is to be expected since loop volume increases with aspect ratio, whereas the size of the stability region shrinks, see Figure \ref{its}. Secondly, as indicated before (Figure \ref{it_a1p_ar20_ax_wr}, right), increasing the relaxation radius increases the energy released. And thirdly, correlated walks mean higher step counts, however, whether or not there is also an increased energy release depends on the relaxation radius.

If \textit{R}$_x$\,=\,\textit{R}$_3$(=\,1), the energy release from correlated walks is reduced compared to the uncorrelated distributions; whereas, complete relaxation, \textit{R}$_x$\,=\,\textit{R}$_4$(=\,3), leads to an increased energy release. This less-than-straightforward point is consistent with the plot that shows the variation in energy release along the threshold for both values of \textit{R}$_x$, see Figure \ref{it_a1p_ar20_ax_wr} (right). A correlated walk would favour crossings around threshold point 90; when \textit{R}$_x$\,=\,\textit{R}$_3$ the energy release is almost at its lowest for this part of the threshold, whereas the opposite is the case when \textit{R}$_x$\,=\,\textit{R}$_4$. This is also true for the thresholds applicable to loops of aspect ratio 10 and 30.

For loops of aspect ratio 10, correlated walks produce distributions that have high-energy cut-offs - this feature is an artifact of the simple two-$\alpha$ model. It is caused by the fact that when \textit{L}\,/\textit{R}$_{3}$\,=\,10, the relaxation line intersects the \textit{B}$_z$ reversal line before the instability threshold.

When one calculates the dimensional heat fluxes (Equation \ref{dim_energy_flux}) one finds that flux is weakly dependent on aspect ratio. Further examination reveals that any dependence on aspect ratio can only come from $\sum\delta W^{\hspace{0.02cm}*}$, which is determined by the coordinates of the instability threshold. $\delta W^{\hspace{0.02cm}*}$ incorporates a length factor in units of the loop radius, i.e., $(L/R_c)\,\delta w^{\hspace{0.02cm}*}=\delta W^{\hspace{0.02cm}*}$, where $\delta w^{\hspace{0.02cm}*}$ is the dimensionless energy release per unit of dimensionless length. Substituting the full expression for the step time ($\tau\,=\,(\lambda/2)(L/v_{\theta})(R_f/R_c)$, Section \ref{sec:flare_energy_distribution}) into Equation \ref{dim_energy_flux} gives
\begin{eqnarray}
  \label{dim_energy_flux_2}	
  F & = & \frac{81\pi}{10^5\mu_0}\frac{R_c}{R_f}\frac{1}{N\lambda} \hspace{0.1cm} B_c^2 v_{\theta} \hspace{0.1cm} \sum_{i=1}^{10^5}\delta w^{\hspace{0.02cm}*};
\end{eqnarray}
again, applying previously used values, this simplifies to $F\,=\,(10^6/N)\sum\delta w^{\hspace{0.02cm}*}$ erg cm$^{-2}$ s$^{-1}$. The length terms cancel and the ratio $R_c\,/R_f$ is effectively a constant.

For distributions derived from uncorrelated walks and minimal relaxation (\textit{R}$_x$\,=\,\textit{R}$_3$), \textit{F}\,$\approx$\,3-4$\times$10$^6$ $\mathrm{erg\,\,cm^{-2}\,\,s^{-1}}$. Using correlated walks instead, diminishes the fluxes to 0.9-2$\times$10$^6$ $\mathrm{erg\,\,cm^{-2}\,\,s^{-1}}$. Increasing the relaxation radius to \textit{R}$_4$ will reverse this reduction and yield \textit{F}\,$\approx$\,7-10$\times$10$^6$ $\mathrm{erg\,\,cm^{-2}\,\,s^{-1}}$. This last result is also true for distributions based on uncorrelated walks and full relaxations. When \textit{R}$_x$\,=\,\textit{R}$_4$ correlated walks do lead to higher energy releases, however, these walks are longer and have higher step counts, which means the flux remains roughly constant.

\begin{figure}[h!] 
  \center
  \includegraphics[scale=0.44]{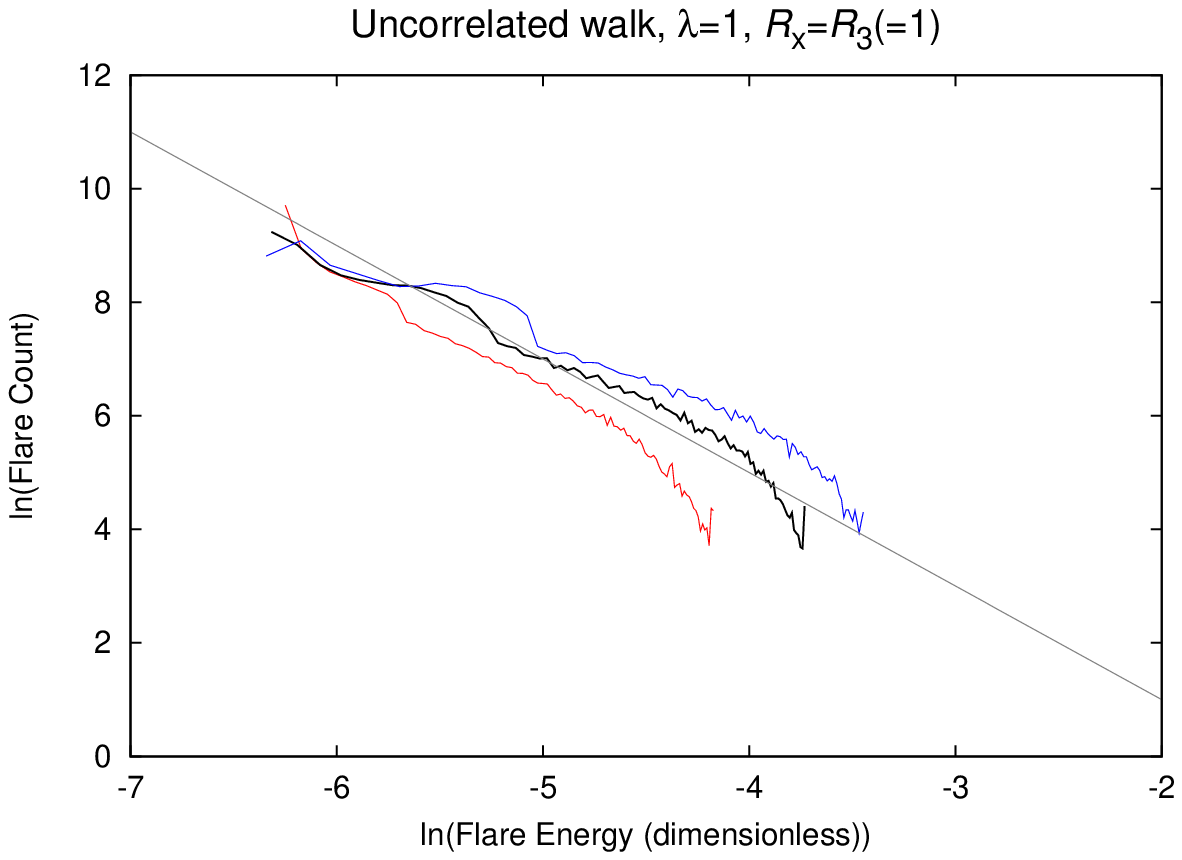}
  \includegraphics[scale=0.44]{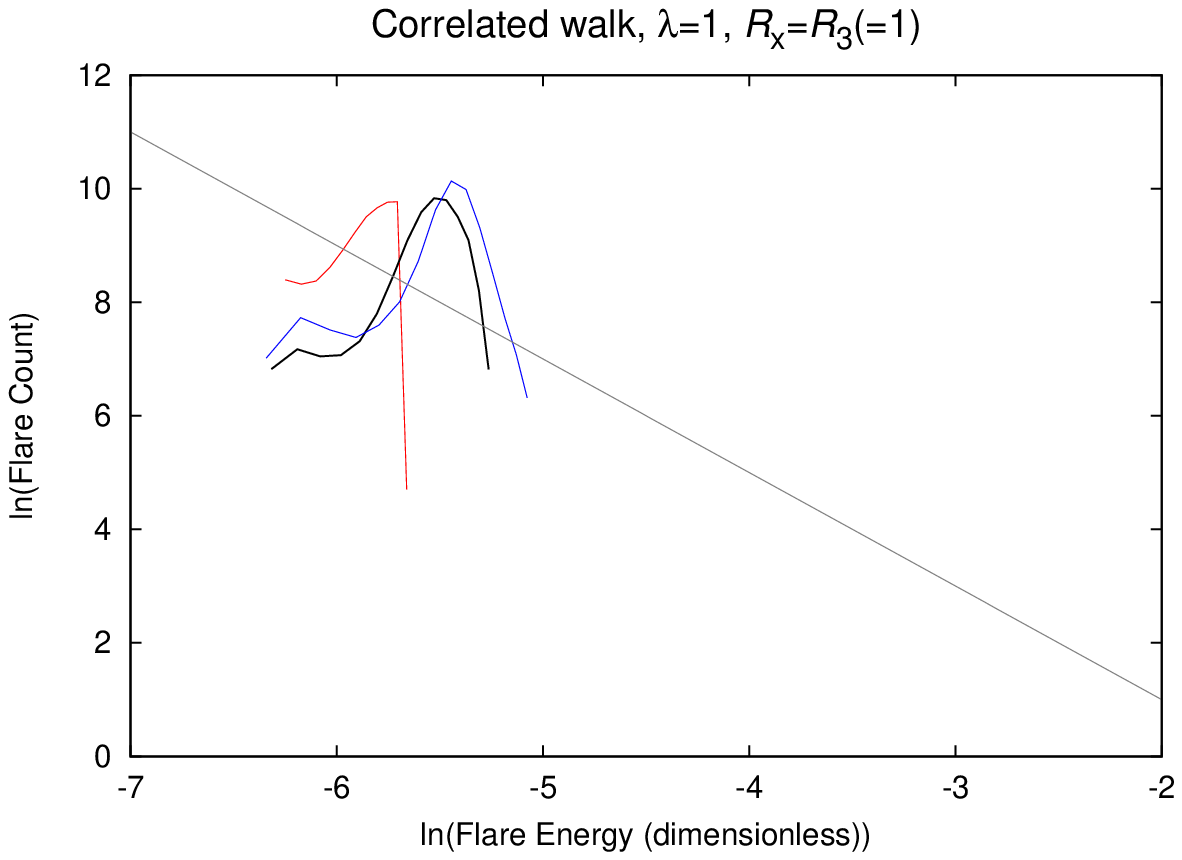}
  \includegraphics[scale=0.44]{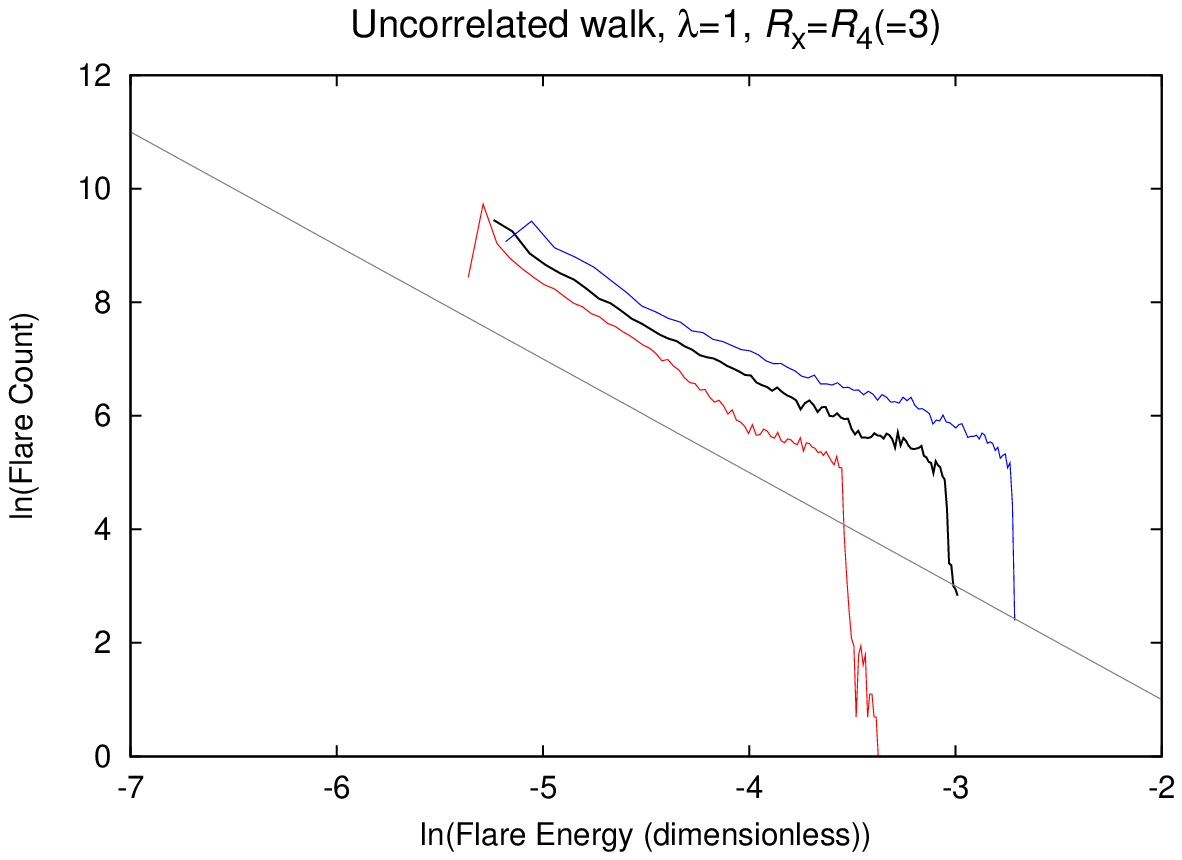}
  \includegraphics[scale=0.44]{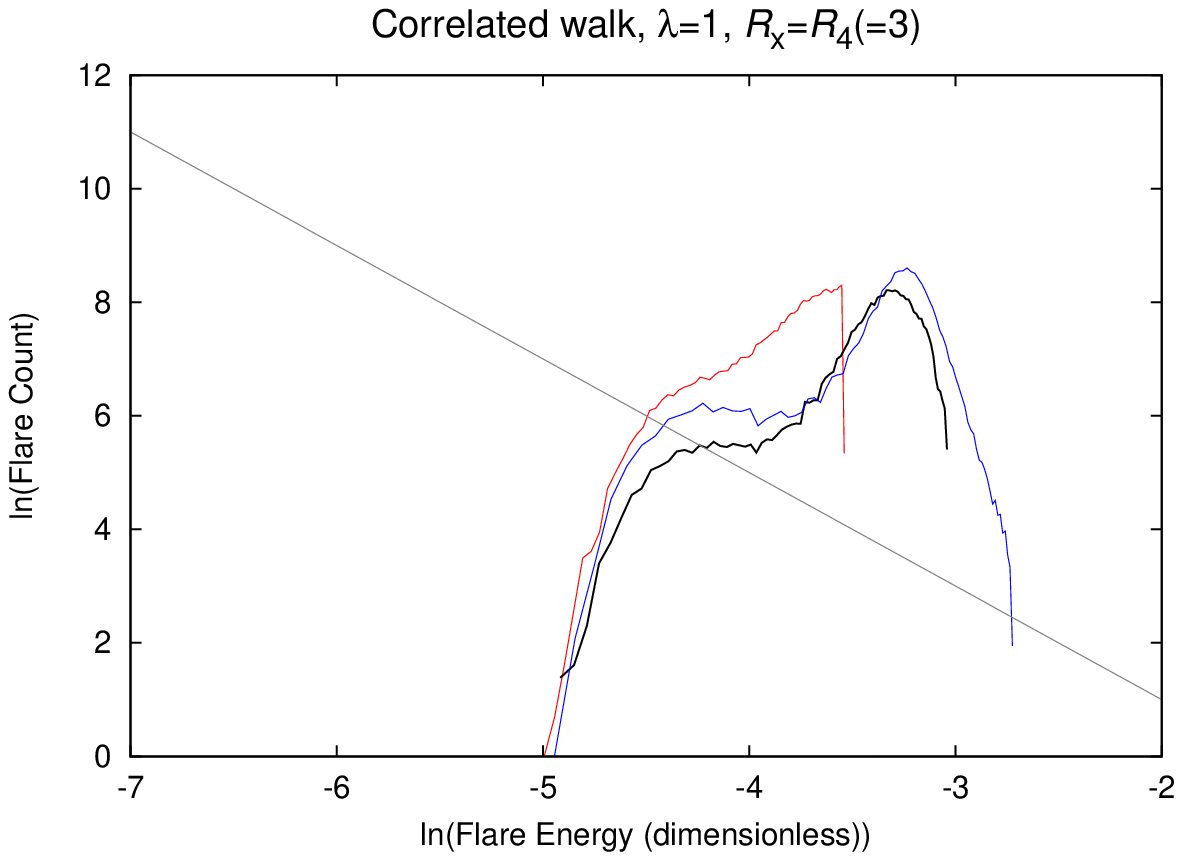}
  \caption{\small{The logarithm of the flare energy distributions for a 10$^5$ loop ensemble, with each loop undergoing one relaxation event. The presentation of these plots follows the scheme used for Figure \ref{wrpf}. The grey diagonal line in each plot is there for comparison; it has a gradient equal to the critical gradient for coronal heating, -2.}}
  \label{wrpl}
\end{figure}

Finally, in Figure \ref{wrpl} we show the \textit{logs} of the flare energy distributions presented in Figure \ref{wrpf}. The distributions calculated from uncorrelated walks give log plots that almost match the critical gradient for coronal heating. Although the log plots of the correlated (Gaussian-shaped) distributions do not follow power-laws, we include these results for completeness.

\section{Summary and conclusions}
We have investigated the distribution of energy releases in an ensemble of coronal loops driven by random photospheric footpoint motions, using Taylor relaxation theory. The twisting has been assumed to be localised within the loop cross section, so that the loop is always without net current (the azimuthal field vanishes at - and beyond - the loop boundary). This means we are genuinely studying individual loops, rather than (unrealistically) allowing the potential envelope outside the loop to be twisted as the loop evolves, as in previous work (BBV\citeyear{Bareford2010}).

A relaxation event is triggered whenever the loop becomes unstable to an ideal kink instability; during the nonlinear phase, current sheets form and subsequent rapid reconnection occurs. A distribution of events is built up by allowing loop equilibria to evolve through a random walk, representing the effects of turbulent photospheric footpoint motions, until the linear stability threshold is reached.  The effectiveness of the relaxation approach is that energy release is easily  calculated for a wide family of profiles, this is extremely difficult with 3D numerical simulations. Furthermore, relaxation theory becomes a better representation for very high conductivities, which cannot be accessed by present day simulations. This approach can, of course, be extended to more complex field models than the simple cylindrical coronal loop models used here. The energy fluxes obtained are sufficient for coronal heating; the fluxes agree with the oft-quoted estimates of Withbroe and Noyes (\citeyear{Withbroe1977}). The heating flux is larger for photospheric motions with higher temporal correlation: within our model, this is represented by larger step sizes for the random walk (so that the loop is coherently twisted until it reaches instability, rather than randomly twisting and untwisting many times). We thus show that dissipation within loops could be sufficient for coronal heating, but in reality, this is likely to form only part of the coronal heating input. Topological complexity arising from, for example, the discrete nature of photospheric flux sources (Priest, Heyvaerts, and Title, \citeyear{Priest2002}) and braiding motions (Parker, \citeyear{Parker1972}), will also play a strong role.

A distribution of heating events, or nanoflares, is obtained, for a variety of conditions. For the case of spatially uncorrelated twisting motions, in which the motions may vary strongly across the loop cross section, a power law distribution of energy versus occurrence frequency is obtained, with a slope slightly steeper than the critical value of -2 required for nanoflare heating to be effective (Hudson, \citeyear{Hudson1991}). For strongly correlated twist motions, in which the twist in the outer part of the loop is close to that in the inner core, a peaked energy distribution is obtained, with almost Gaussian shape. The former case reflects the distribution of available energies around the instability threshold, whereas as the latter is mainly determined by the allowable range of twist profiles. It should be noted that these distributions (Figures \ref{wrpf}-\ref{wrpl}) are obtained for an ensemble of identical loops: in reality, much broader distributions will result due to variations in axial field strengths and photospheric driving. The true nanoflare distribution is a convolution over more than one parameter.

The effect of loop aspect ratio has been considered and been found to have little impact on energy flux. The higher volume of large aspect ratio loops is counteracted by the smaller stability region (instability occurs at lower $\alpha$ values). As the aspect ratio is increased beyond 30, we expect the stability region to reduce by smaller and smaller amounts. In other words, the region will converge to a minimum area. This has been shown for constant-$\alpha$ loops, see Figure 4 of Browning and Van der Linden (\citeyear{Browning2003}). Hence, assuming this expectation is verified, the energy flux will be independent of aspect ratio (Equation \ref{dim_energy_flux_2}), assuming that the same axial field strength is applied to all members of the loop ensemble. Presumably, there is a dependence between loop size and $|\textit{B}_c|$, so an ensemble that features some distribution of field strengths will still depend (albeit indirectly) on the aspect ratio.

Contrary to the \textit{B}$_z$ profiles of Appendix B, the axial field at the loop footpoints should not change during the random walk or during relaxation. The reason for this discrepancy is that preservation of the footpoint axial field introduces a \textit{z} dependency - the field becomes two dimensional. However, if the length of the loop exceeds its radius, a 1D field approximation - such as the one used by the model presented here - still remains adequate for a substantial portion of the loop. Zweibel and Boozer (\citeyear{Zweibel1985}) and Browning and Hood (\citeyear{Browning1989}) show that the \textit{z} dependence is confined to thin boundary layers near the footpoints. Hence, the difference in energies for loops represented by 1D and 2D fields is negligible especially if \textit{L}\,/\textit{R}$_3$\,$>$\,10 (see also Lothian and Browning, \citeyear{Lothian2000}; Robertson, Hood, and Lothian, \citeyear{Robertson1992}). Dalmasse, Browning and Bareford (\citeyear{Dalmasse2011}) have investigated a simpler loop, having just a core and outer layer (with a conducting wall at \textit{r}\,=\,1), by calculating the energy releases according to Taylor relaxation for a representative sample of threshold configurations. This was done using both 1D and 2D fields, with the latter maintaining the axial field at the footpoints. The resulting energy releases differ by less than 1\% between the 1D and 2D cases.

The results presented here are based on a loop model that has a thin current-neutralising layer (this approximates to a current sheet), in which the fields discontinuously change at the loop edge. The main reason for this choice is so that the fields inside the loop are close to the previously-studied two-$\alpha$ model (BBV\citeyear{Bareford2010}), and thus a comparison can be made with previous work. Also, such fields correspond to twisting within an isolated flux source, whilst the flux which surrounds the loop in the corona originates from untwisted separated sources. Interestingly, the ideal instability threshold in this case is very similar to that found for a close-fitting conducting wall at the loop edge, as originally used by Browning and Van der Linden (\citeyear{Browning2003}). This is because the thin current layer forces unstable perturbations to vanish (almost) at the loop edge. In numerical simulations, the choice of a thin current layer has consequences in allowing resistive modes to be significant; although for realistic values of the resistivity (unattainable in simulations) the growth rate of such modes is extremely slow. Preliminary studies have also been undertaken with a thicker current-neutralising layer. In this case, a closed stability threshold curve can be obtained, and the results are more similar to previous work (BBV\citeyear{Bareford2010}).

One important consequence of considering loops with zero net current is that the reconnection activity tends to be localised near the loop and thus relaxation is likely to be incomplete (rather than including a large part of the surrounding potential field). We consider here two limiting cases: localised relaxation, in which only the loop volume relaxes to a minimum energy (constant-$\alpha$) state, and the surrounding potential envelope remains unaffected; and complete relaxation, in which the loop and the potential envelope relax out to the external boundary. The latter is clearly the true minimum energy state. Numerical simulations (Bareford et al., \citeyear{Bareford2011}) indicate an intermediate situation, but somewhat closer to the completely localised relaxation. In fact, the loop reconnects with some of the surrounding axial field, but only to a limited extent. This is an important issue for understanding relaxation in the Sun, where the extent of relaxation is not defined by conducting walls: in contrast with laboratory plasmas (Taylor, \citeyear{Taylor1974}). This is discussed more fully in a companion paper (Bareford et al., \citeyear{Bareford2011}). In general, complete relaxation naturally gives larger energy releases, but the choice of relaxation radius does not strongly affect the distribution of heating events. Future work will use numerical simulations to explore the transition to instability, and the effects of continual driving.
\\\\
\small{\textit{Acknowledgements}. The authors thank Jim Klimchuk for constructive comments and M. R. B. thanks STFC for studentship support.}

\appendix
\section{Expressions for loop properties}
Expressions for some key quantities ($\langle\tilde{\varphi}\rangle$, \textit{K} and \textit{W}) are given here. For compactness, these are given only for $\alpha_{1}$\,$\neq$\,0 and $\alpha_{2}$\,$\neq$\,0, while special cases (e.g., $\alpha_1$\,=\,0) must be dealt with separately. Expressions for constant-$\alpha$ fields can be recovered by setting $\alpha_1$\,=\,$\alpha_2$, which gives more familiar formulae. The superscripts and subscripts that accompany each quantity term denote the upper and lower radial bounds over which the quantity is calculated. The vacuum permeability, $\mu_0$, used in the magnetic energy expressions, is set to 1. 

\subsection{Average magnetic twist}
\begin{eqnarray}
  \langle\tilde{\varphi}\rangle_{R_{0}}^{R_{1}} & = & \frac{\sigma_1 L\Big[1-J_{0}(|\alpha_{1}|R_{1})\Big]}{R_1 J_{1}(|\alpha_{1}|R_{1})}\\
  \nonumber   &   &\\  
  \langle\tilde{\varphi}\rangle_{R_{1}}^{R_{2}} & = & \frac{\sigma_2 L\Big[F_0(|\alpha_2|R_1)-F_0(|\alpha_2|R_2)\Big]}{R_2 F_1(|\alpha_2|R_2)-R_1 F_1(|\alpha_2|R_1)}\\
  \nonumber   &   &\\
  \langle\tilde{\varphi}\rangle_{R_{2}}^{R_{3}} & = & \frac{\sigma_3 L R_2\Big[G_0(|\alpha_3|R_2) + G_0(|\alpha_3|R_3)\Big]}{R_3 G_1(|\alpha_3|R_3) + R_2 G_1(|\alpha_3|R_2)}\\ 
  \nonumber   &   &\\
  \langle\tilde{\varphi}\rangle_{R_{3}}^{R_{4}} & = & 0 
\end{eqnarray}
\newpage
\subsection{Magnetic helicity}
\begin{eqnarray}  
  \nonumber K_{R_{0}}^{R_{1}} & = & \sigma_1\frac{2\pi L B_1^2}{|\alpha_1|}\Bigg(R_1^2 J_0^2(|\alpha_1|R_1)+R_1^2 J_1^2(|\alpha_1|R_1)-2\frac{R_1}{|\alpha_1|}J_0(|\alpha_1|R_1)J_1(|\alpha_1|R_1)\Bigg)\\
\end{eqnarray}

\begin{eqnarray}  
  \nonumber K_{R_{1}}^{R_{2}} & = & \sigma_2\frac{2\pi L B_2^2}{|\alpha_2|}\Bigg(R_2^2 F_0^2(|\alpha_2|R_2)+R_2^2 F_1^2(|\alpha_2|R_2)-2\frac{R_2}{|\alpha_2|}F_0(|\alpha_2|R_2)F_1(|\alpha_2|R_2)\Bigg)\\
  \nonumber   &   &\\
  \nonumber &   &\,-\,\sigma_2\frac{2\pi L B_2^2}{|\alpha_2|}\Bigg(R_1^2 F_0^2(|\alpha_2|R_1)+R_1^2 F_1^2(|\alpha_2|R_1)-2\frac{R_1}{|\alpha_2|}F_0(|\alpha_2|R_1)F_1(|\alpha_2|R_1)\Bigg)\\
  \nonumber   &   &\\
  \nonumber   &   & \,+\,\sigma_2\frac{4\pi L B_2}{|\alpha_2|}\Big(F_0(|\alpha_2|R_1) - F_0(|\alpha_2|R_2)\Big)\Bigg[B_1 R_1 J_1(|\alpha_1|R_1) \Bigg(\frac{1}{|\alpha_1|} - \frac{\sigma_{1,2}}{|\alpha_2|}\Bigg)\Bigg]\\ 
  &   &\\
  \nonumber   &   &\\
  \nonumber K_{R_{2}}^{R_{3}} & = & \sigma_3\frac{2\pi L B_3^2}{|\alpha_3|}\Bigg(R_3^2 G_0^2(|\alpha_3|R_3)+R_3^2 G_1^2(|\alpha_3|R_3)-2\frac{R_3}{|\alpha_3|}G_0(|\alpha_3|R_3)G_1(|\alpha_3|R_3)\Bigg)\\
  \nonumber   &   &\\
  \nonumber &   &\,-\,\sigma_3\frac{2\pi L B_3^2}{|\alpha_3|}\Bigg(R_2^2 G_0^2(|\alpha_3|R_2)+R_2^2 G_1^2(|\alpha_3|R_2)-2\frac{R_2}{|\alpha_3|}G_0(|\alpha_3|R_2)G_1(|\alpha_3|R_2)\Bigg)\\
  \nonumber   &   &\\
  \nonumber   &   & \,+\,\sigma_3\frac{4\pi L B_3}{|\alpha_3|}\Big(G_0(|\alpha_3|R_2) - G_0(|\alpha_3|R_3)\Big)\Bigg[B_2 R_2 F_1(|\alpha_2|R_2) \Bigg(\frac{1}{|\alpha_2|} - \frac{\sigma_{2,3}}{|\alpha_3|}\Bigg)\\
  &   & \,\,\,\,\,\,\,\,\,\,\,\,\,\,\,\,\,\,\,\,\,\,\,\,\,\,\,\,\,\,\,\,\,\,\,\,\,\,\,\,\,\,\,\,\,\,\,\,\,\,\,\,\,\,\,\,\,\,\,\,\,\,\,\,\,\,\,\,\,\,\,\,\,\,\,\,\,\,\,\,\,\,\,\,\,\,\,\,\,\,\,\, + \,B_1 R_1 J_1(|\alpha_1|R_1) \Bigg(\frac{1}{|\alpha_1|} - \frac{\sigma_{1,2}}{|\alpha_2|}\Bigg)\Bigg]\\
  \nonumber   &   &\\  
  \nonumber   &   &\\  
  K_{R_{3}}^{R_{4}} & = & 0  
\end{eqnarray}

\subsection{Magnetic energy}
\begin{eqnarray}
  \nonumber W_{R_{0}}^{R_{1}} & = & \frac {L\pi B_1^2}{\mu_0}\Bigg[R_1^2\Big(J_0^2(|\alpha_1|R_1) + J_1^2(|\alpha_1|R_1)\Big) - \frac{R_1}{|\alpha_1|}J_0(|\alpha_1|R_1)J_1(|\alpha_1|R_1)\Bigg]\\
  &   &\\
  \nonumber &   &\\
  \nonumber &   &\\  
  \nonumber W_{R_{1}}^{R_{2}} & = & \frac {L\pi B_2^2}{\mu_0}\Bigg[R_2^2\Big(F_0^2(|\alpha_2|R_2) + F_1^2(|\alpha_2|R_2)\Big) - \frac{R_2}{|\alpha_2|}F_0(|\alpha_2|R_2)F_1(|\alpha_2|R_2)\\
  \nonumber &   & \,\,\,\,\,\,\,\,\,\,\,\,\,\,\,\,\,\,-\,R_1^2\Big(F_0^2(|\alpha_2|R_1) + F_1^2(|\alpha_2|R_1)\Big) + \frac{R_1}{|\alpha_2|}F_0(|\alpha_2|R_1)F_1(|\alpha_2|R_1)\Bigg]\\
  &   &\\
  \nonumber &   &\\
  \nonumber W_{R_{2}}^{R_{3}} & = & \frac {L\pi B_3^2}{\mu_0}\Bigg[R_3^2\Big(G_0^2(|\alpha_3|R_3) + G_1^2(|\alpha_3|R_3)\Big) - \frac{R_3}{|\alpha_3|}G_0(|\alpha_3|R_3)G_1(|\alpha_3|R_3)\\
  \nonumber &   & \,\,\,\,\,\,\,\,\,\,\,\,\,\,\,\,\,\,-\,R_2^2\Big(G_0^2(|\alpha_3|R_2) + G_1^2(|\alpha_3|R_2)\Big) + \frac{R_2}{|\alpha_3|}G_0(|\alpha_3|R_2)G_1(|\alpha_3|R_2)\Bigg]\\
  &   &\\
  \nonumber &   &\\ 
  W_{R_{3}}^{R_{4}} & = & \frac{L\pi}{\mu_0}\Bigg[\frac{B_4^2}{2}\Big(R_4^2 - R_3^2\Big)\Bigg]    
\end{eqnarray}

\section{Magnetic field profiles for a selection of $\alpha$-space points}

\begin{figure}[h!]
  \center
  \includegraphics[scale=0.6]{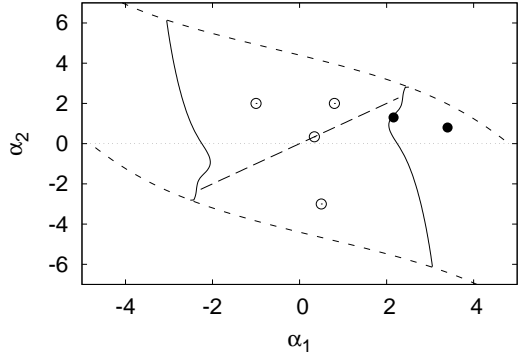}
  \caption{\small{The stability region for a loop of aspect ratio 20 is demarcated by instability thresholds (solid lines) and \textit{B}$_z$ reversal lines (short dashed lines). The relaxation line (long dashed) comprises the points within the stability region where $\alpha_1$\,=\,$\alpha_2$. Stable configurations are indicated by empty circles and unstable ones by filled circles.}}
  \label{it_ar20_xpts}  
\end{figure}

The magnetic axial field, \textit{B}$_z(\textit{r})$, azimuthal field, \textit{B}$_{\theta}(\textit{r})$, and magnetic twist, $\phi$(\textit{r}), profiles are presented for a selection of stable and unstable loop configurations, see Fig. \ref{it_ar20_xpts}.

\begin{figure}[h!]
  \center
  \includegraphics[scale=0.4]{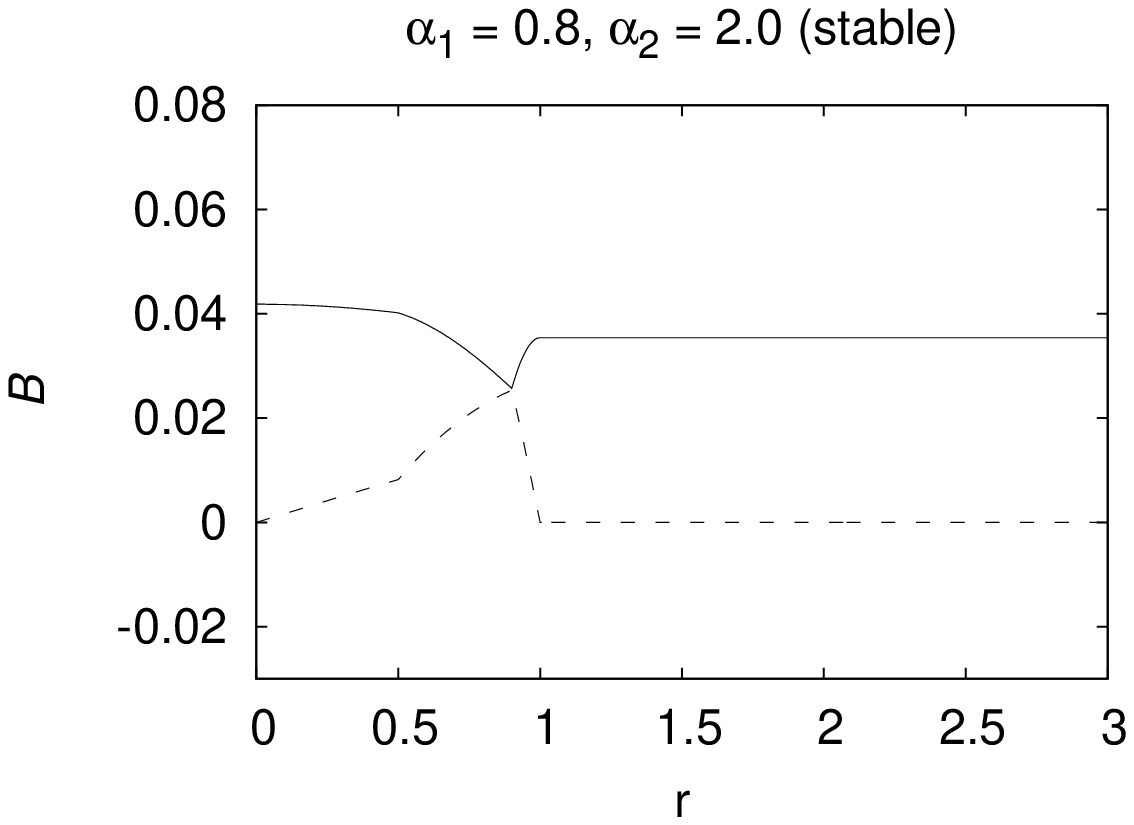}
  \includegraphics[scale=0.4]{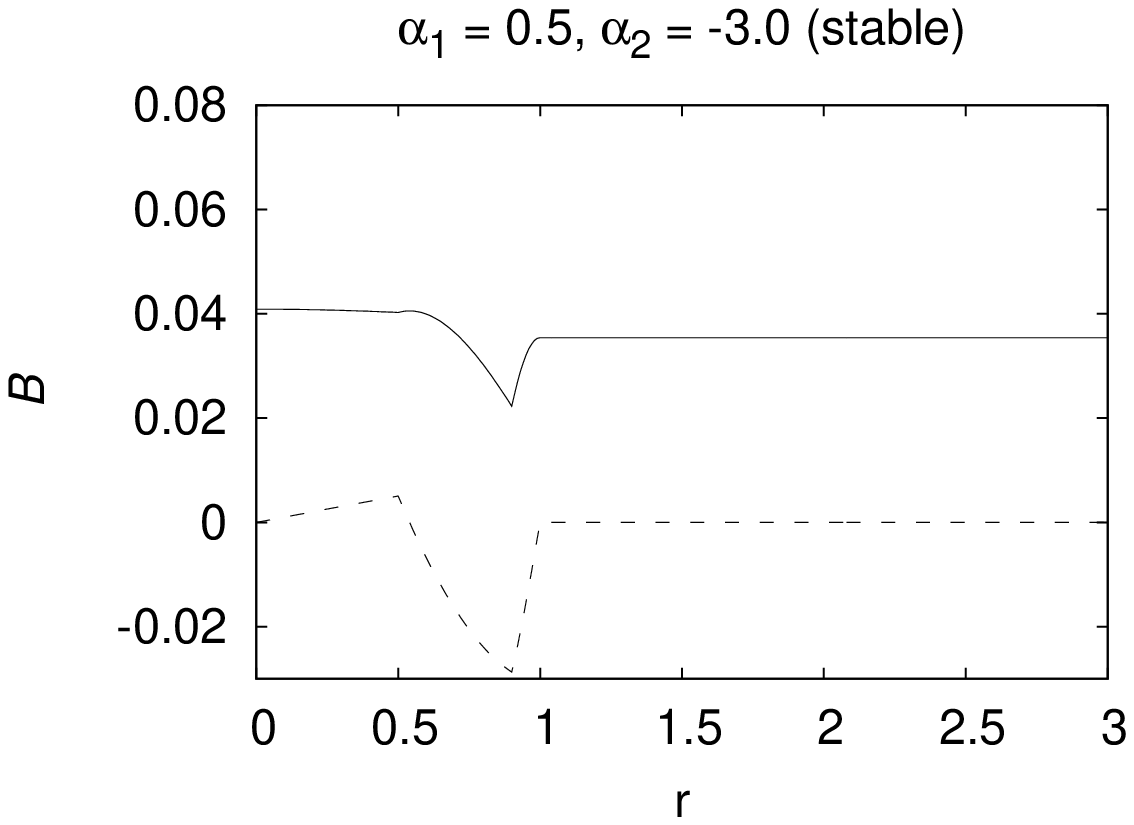}
  \includegraphics[scale=0.4]{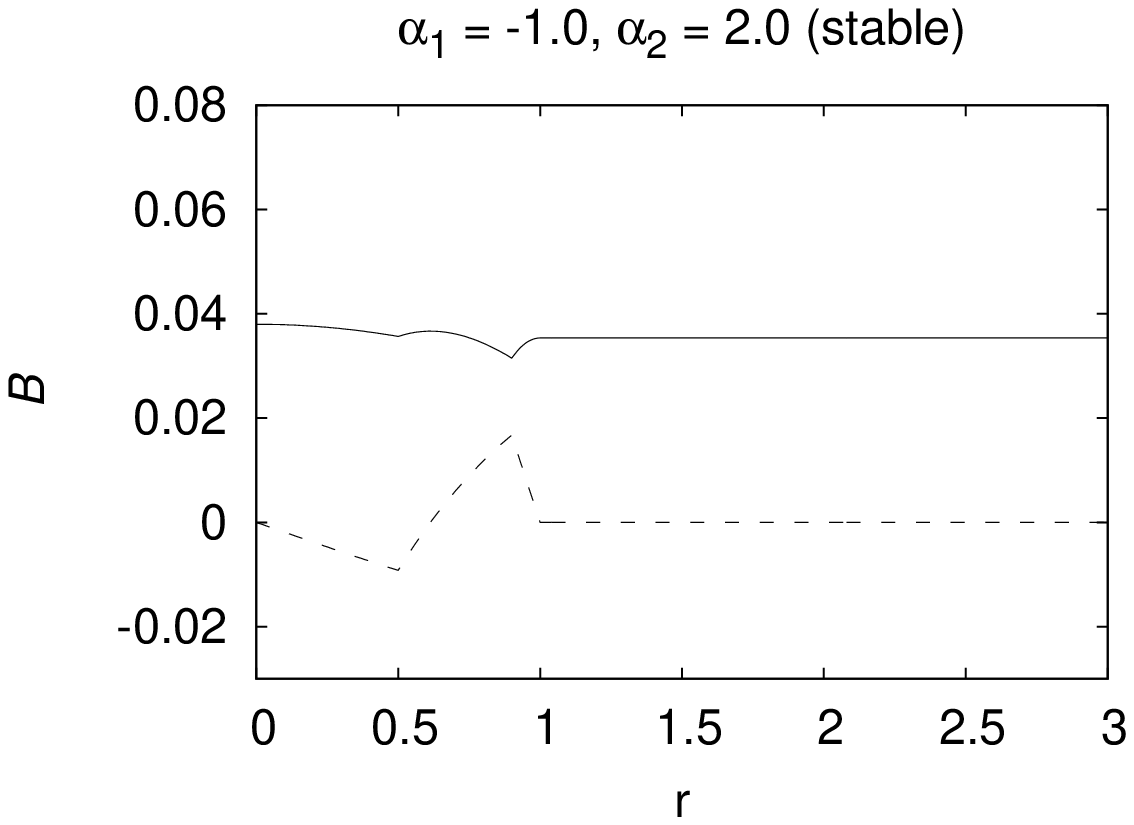}
  \includegraphics[scale=0.4]{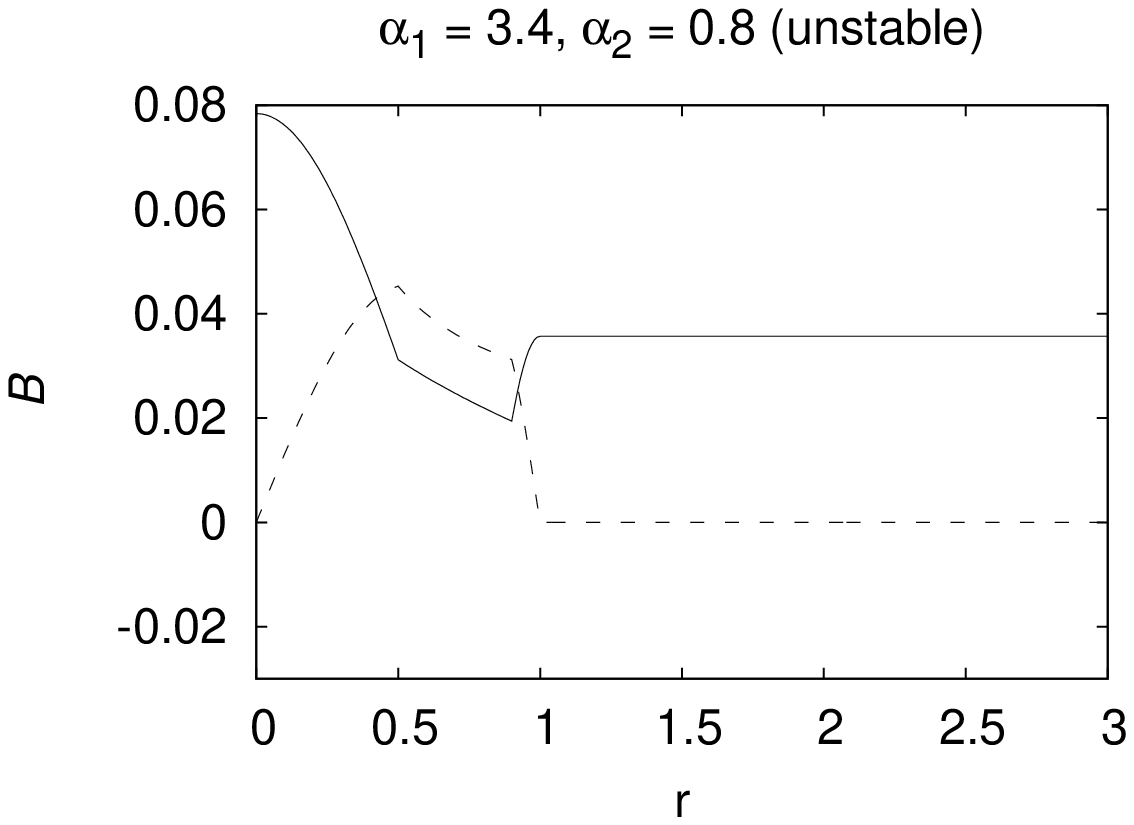}
  \caption{\small{The \textit{B}$_z$ (solid) and \textit{B}$_{\theta}$ (dashed) profiles for some of the configurations (3 stable, 1 unstable) located on Figure \ref{it_ar20_xpts}.}}
  \label{bz_pf}  
\end{figure}

\newpage
\begin{figure}[h!]
  \center
  \includegraphics[scale=0.4]{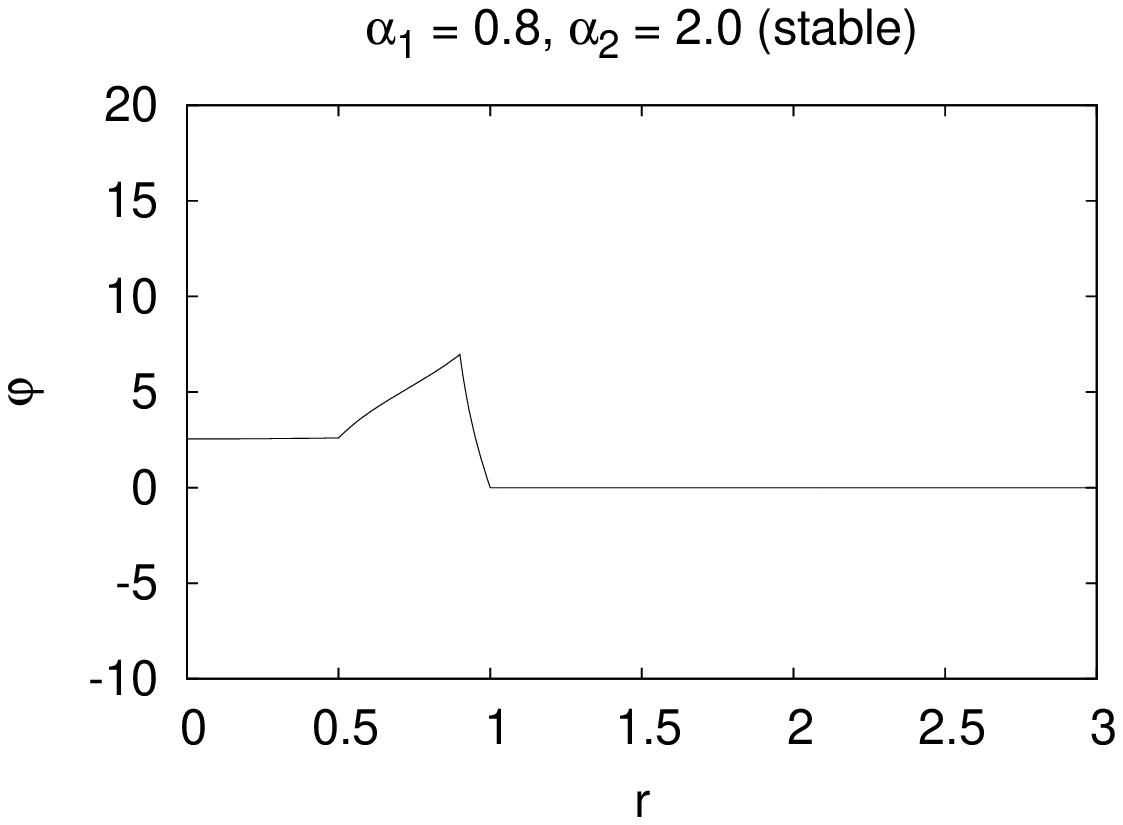}
  \includegraphics[scale=0.4]{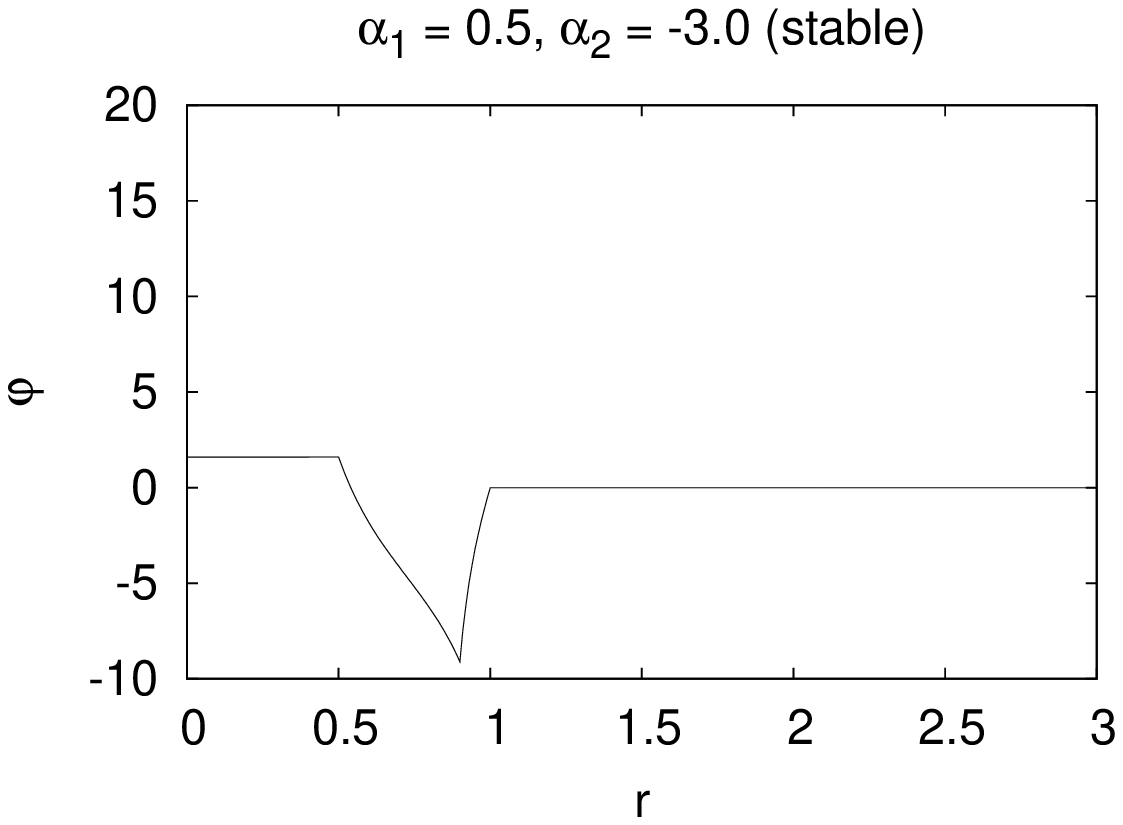}
  \includegraphics[scale=0.4]{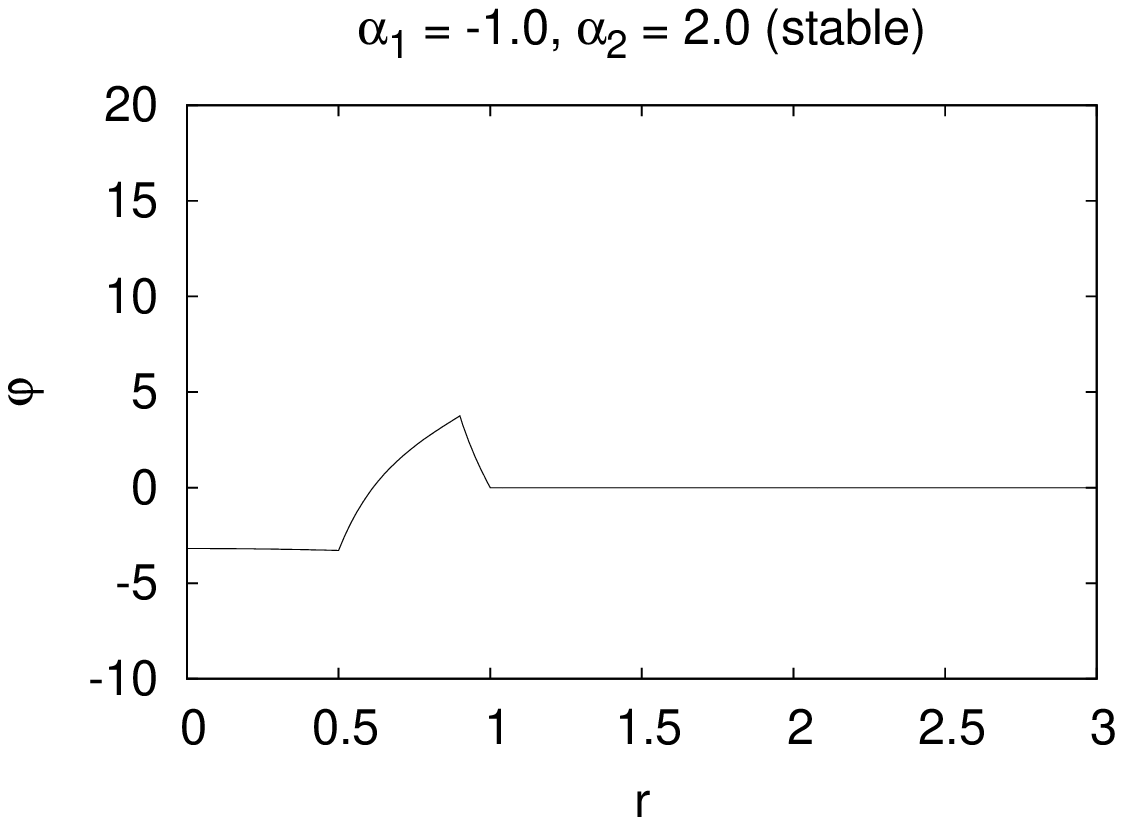}
  \includegraphics[scale=0.4]{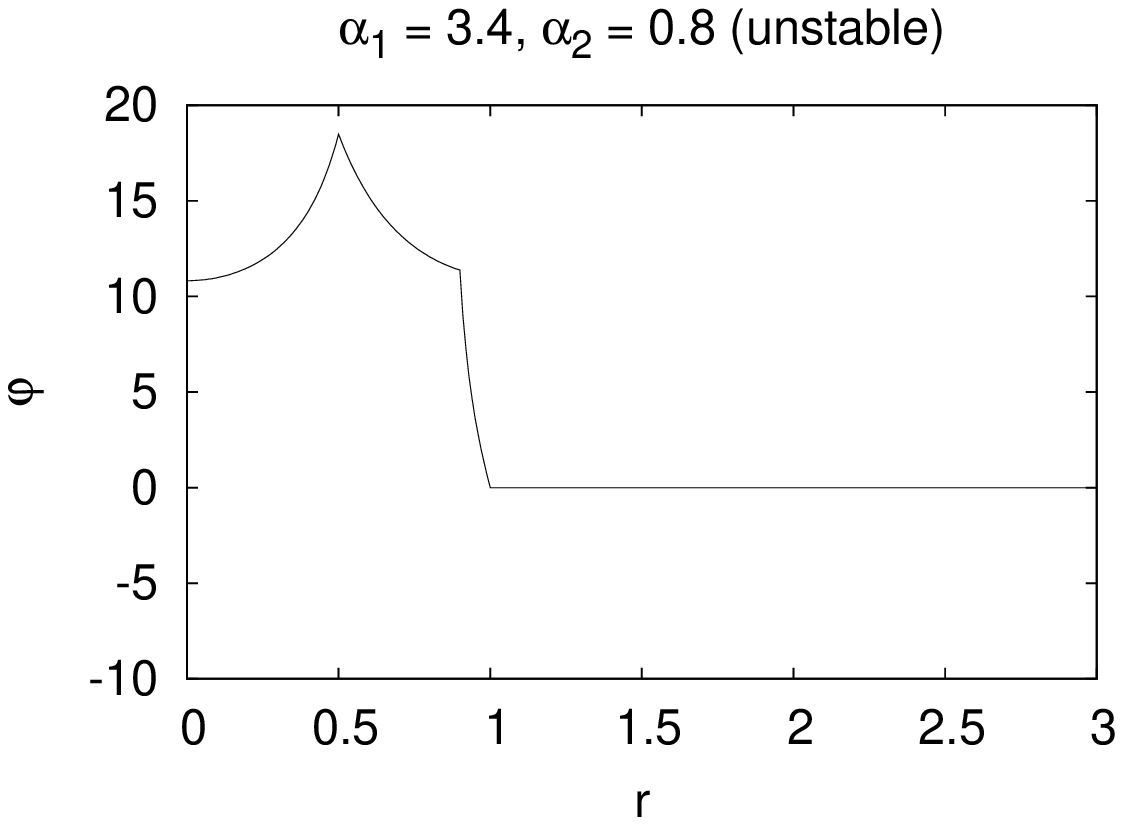}
  \caption{\small{The magnetic twist (in units of $\pi$) profiles for some of the configurations (3 stable, 1 unstable) located on Figure \ref{it_ar20_xpts}.}}
  \label{bz_pf}  
\end{figure}

The empty circle on the $\alpha_1$\,=\,$\alpha_2$ line in Fig. \ref{it_ar20_xpts} is the relaxed state of the unstable configuration identified by the filled circle on the threshold. The radius of the relaxed state is 1.5.

\begin{figure}[h!]
  \center
  \includegraphics[scale=0.4]{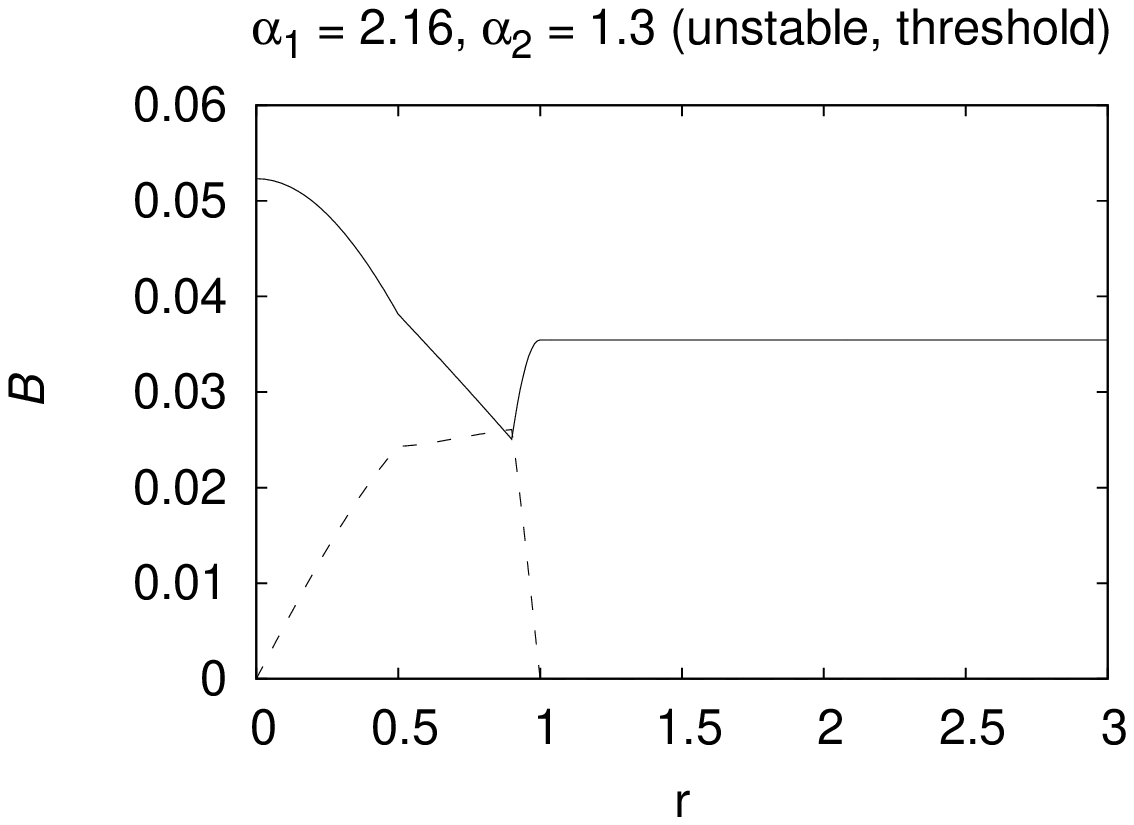}
  \includegraphics[scale=0.4]{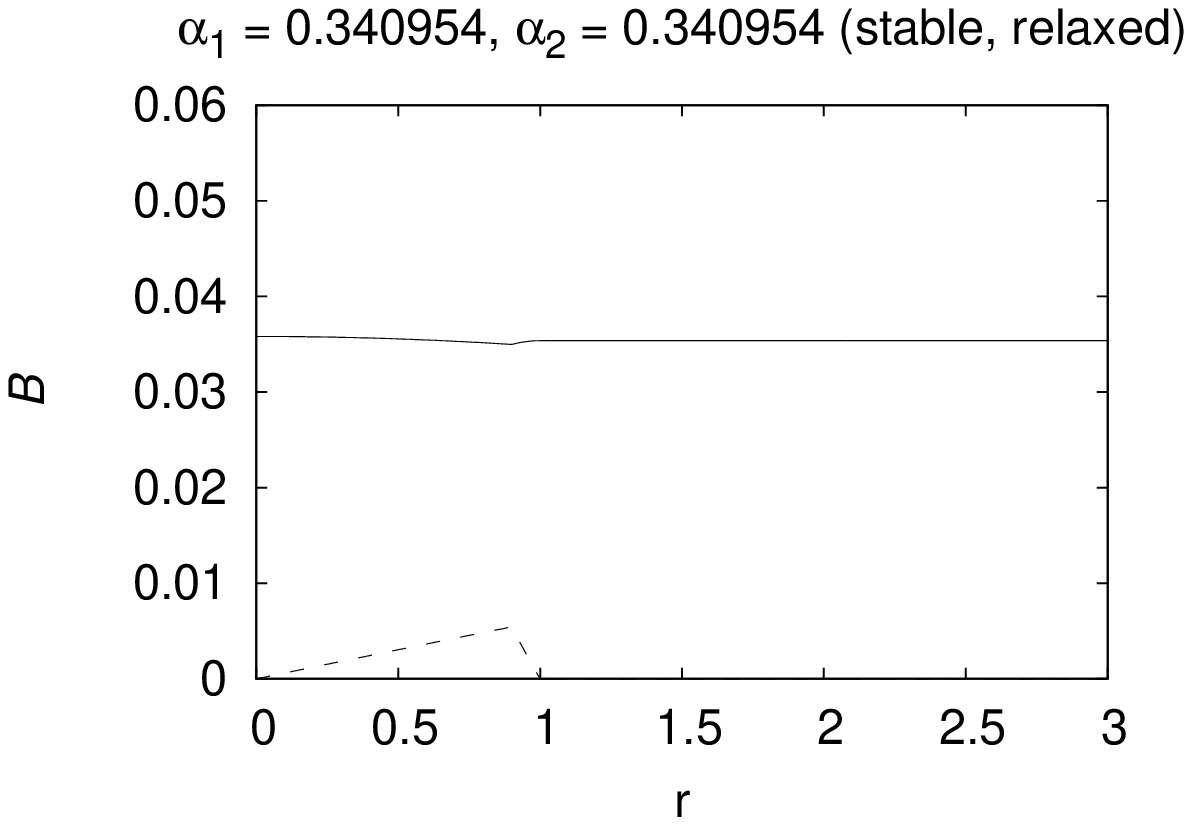}
  \caption{\small{The \textit{B}$_z$ (solid) and \textit{B}$_{\theta}$ (dashed) profiles for the threshold and relaxed configurations located on Figure \ref{it_ar20_xpts}.}}
  \label{bz_pf}  
\end{figure}

\begin{figure}[h!]
  \center
  \includegraphics[scale=0.4]{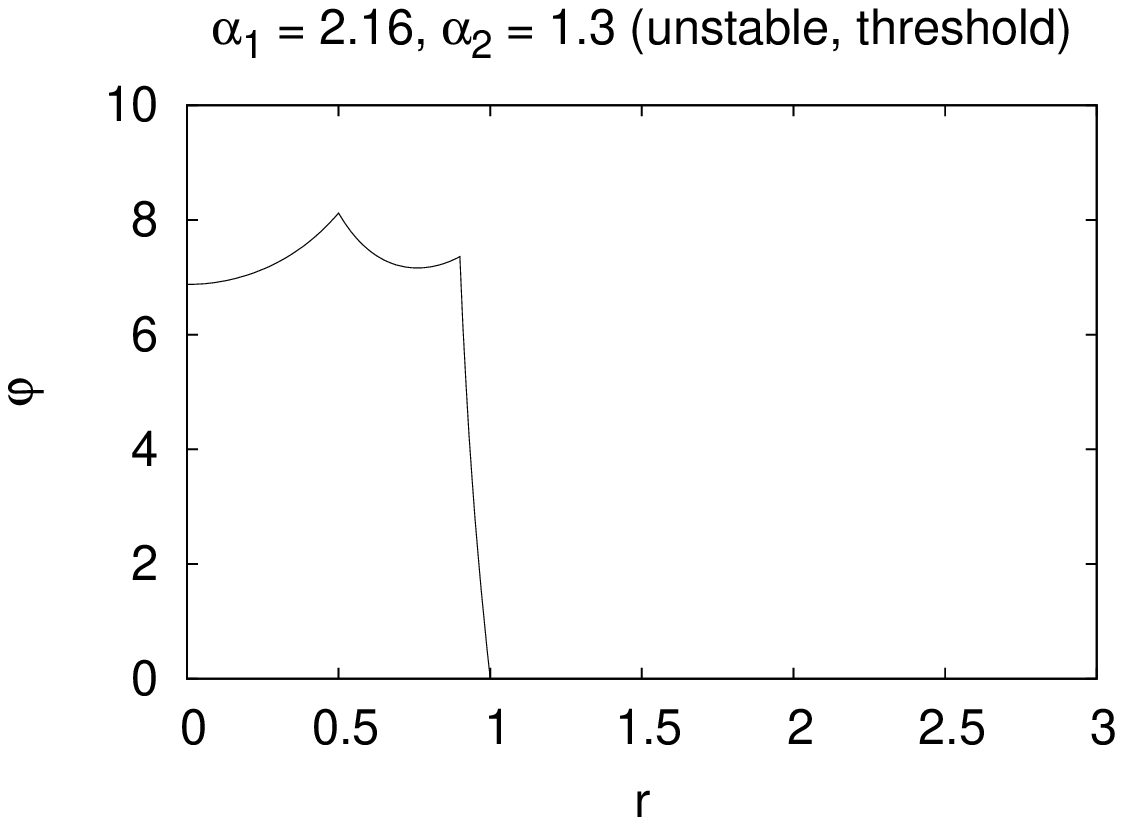}
  \includegraphics[scale=0.4]{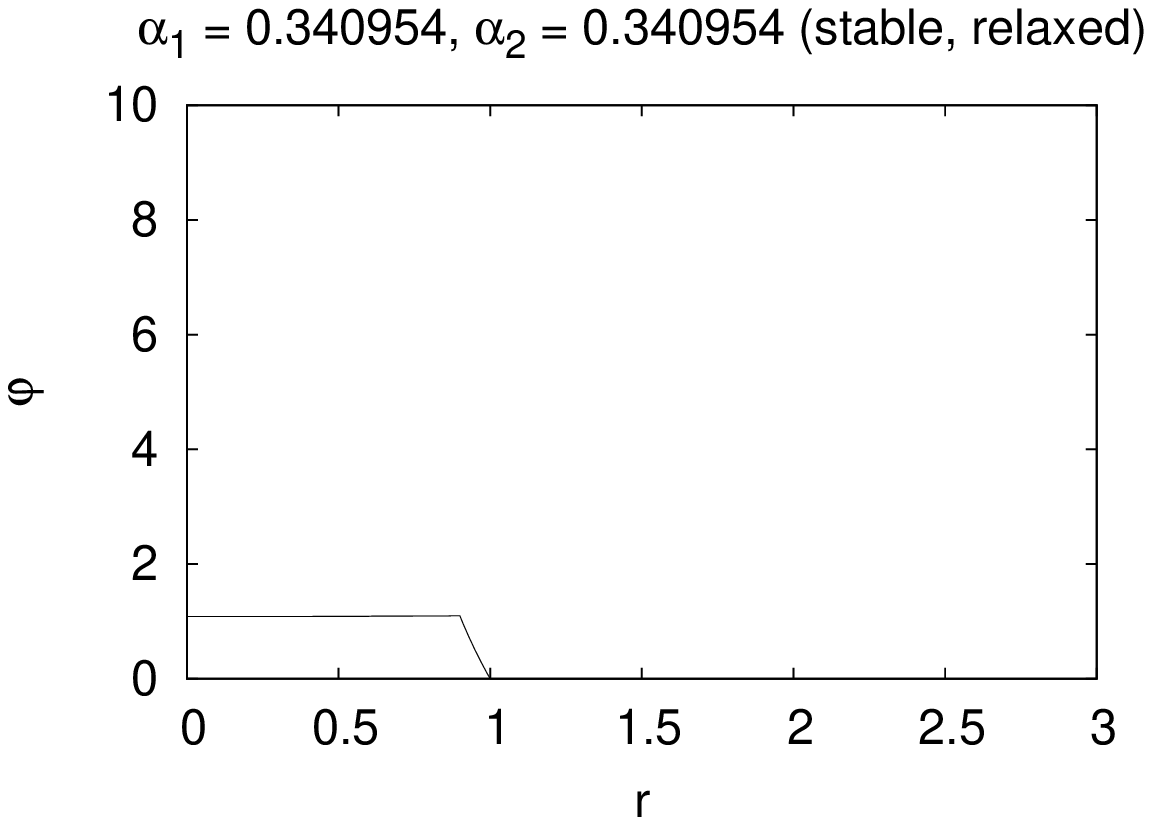}
  \caption{\small{The magnetic twist (in units of $\pi$) profiles for the threshold and relaxed configurations located on Figure \ref{it_ar20_xpts}.}}
  \label{bz_pf}  
\end{figure}

\end{article} 
\end{document}